\documentclass[vecphys]{svmult}

\usepackage{makeidx}         
\usepackage{graphicx}        
\usepackage{multicol}        
\usepackage[bottom]{footmisc}

\makeindex             

\begin{document}

\title*{Precursors and prediction of catastrophic avalanches}
\author{Srutarshi Pradhan\inst{1}\and
Bikas K. Chakrabarti\inst{2}}
\institute{ Norwegian Universty of Science and Technology, Trondheim,
Norway
\texttt{pradhan.srutarshi@ntnu.no}
\and Saha Institute of Nuclear Physics, Kolkata, India
 \texttt{bikask.chakrabarti@saha.ac.in}}
%
%
\maketitle

\section{Introduction}
In this chapter we review the precursors\index{precursor} of catastrophic 
avalanches (global failures)
in several failure models, namely (a) Fiber Bundle Model (FBM), (b)
Random Fuse Model (RFM), (c) Sandpile Models and (d) Fractal Overlap
Model. The precursor parameters identified here essentially reflect
the growing correlations within such systems as they approach their
respective failure points. As we show, often they help us to predict
the global failure points in advance.

Needless to mention that the existence of any such precursors and
detailed knowledge about their behavior for major catastrophic failures 
\cite{pc-books}
like earthquakes\index{earthquake}, landslides, mine/bridge collapses, would be of supreme
value for our civilization. So far, we do not have any established
set of models for these major failure phenomena. However, 
several reasonable models of failure
have already been developed in various contexts. We review
here some of the precursors of global failures in these models. 
\section{Precursors in Failure models }
\subsection{Composite material under stress: Fiber
bundle model }
\begin{center}{\includegraphics[%
  width=2.0in,
  height=1.5in]{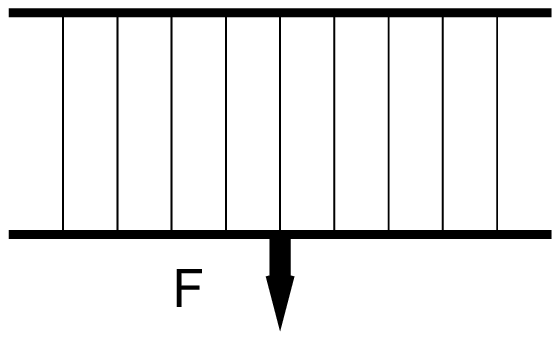}\hskip.6in\includegraphics[%
  width=2.0in,
  height=1.5in]{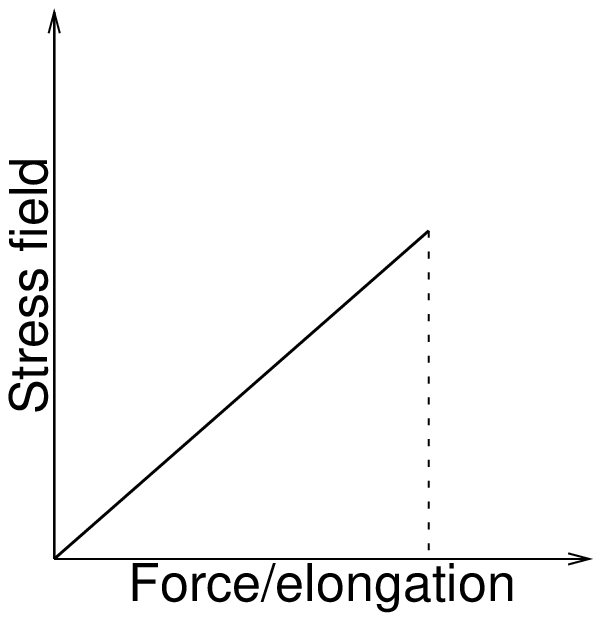}}\end{center}

{\small Fig. 1: Fiber bundle model (left) and the force-response
of a single fiber (right). }{\small \par}

\vskip.1in

Fiber bundle model\index{fiber bundle model} represents various aspects of
fracture-failure process in composite materials. A bundle of parallel fibers,
clamped at both ends (Fig. 1 (left)) represents the model where fibers are 
assumed to obey Hookean elasticity up to the breaking point (Fig. 1 (right)). 
The model needs three
basic ingredients: (a) a discrete set of $N$ elements (b) a probability 
distribution of the strength of fibers (c) a load-transfer rule. 
 Peirce \cite{pc-Peirce} initiated the model study in the context 
of testing the strength
of cotton yarns. Since then this model has been studied from different
views. Fiber bundles are of two classes with respect to the time dependence
of fiber strength: The `static' bundles [2-10] contain fibers
whose strengths are independent of time, where as the `dynamic' bundles
[11-13] are assumed to have time dependent elements to capture
the creep rupture and fatigue behaviors. According to the load sharing
rule,  fiber bundles are being classified into two groups: equal load-sharing\index{equal load sharing}
(ELS) bundles and local load-sharing\index{local load sharing} (LLS). 
In ELS bundles, intact
fibers bear the applied load equally and in LLS bundles, the terminal
load of the failed fiber is given equally to all the intact neighbors.
With steadily increasing load, a fiber bundle approaches the failure
point obeying a dynamics determined by the load sharing rule. The
phase transition \cite{pc-Phase T} and dynamic critical behavior of
the fracture process in such democratic bundles has been established
through recursive formulation \cite{pc-SB01,pc-SBP02} of the failure
dynamics. The exact solutions \cite{pc-SBP02,pc-PSB03} of the recursion
relations suggest universal values of the exponents involved. Attempts
have also been made \cite{pc-variable range} to study the ELS ans LLS
bundles from a common framework introducing a single
parameter which determines the load transfer rule.

We discuss here, failure of  static fiber bundles
under global load sharing (ELS) for two different loading conditions:
(a) load increment by equal amount and (b) quasi-static loading -which
follows weakest link failure at each step of loading. We show analytically
the variation of precursor parameters with the applied stress, which
help to estimate the failure point accurately.

\subsubsection{Recursive dynamics in ELS bundle: Precursors }
ELS model assumes that the intact fibers share the applied load equally.
The strength threshold of a fiber is determined by
the stress value it can bear, and beyond which it fails. Usually,
thresolds are taken from a randomly distributed normalised density
$p(x)$ within the interval $0$ and $1$ such that \begin{equation}
\int_{0}^{1}p(x)dx=1.\label{pc-1}\end{equation}
 The breaking dynamics starts when an initial stress $\sigma$ (load
per fiber) is applied on the bundle. The fibers having strength less
than $\sigma$ fail instantly. Due to this rupture, total number of
intact fibers decreases and effective stress increases and this compels
some more fibers to break. Such stress redistribution and further breaking
of fibers continue till an equilibrium is reached, where either the
surviving fibers are strong enough to bear the applied load on the
bundle or all fibers fail. 

This self organised breaking dynamics can be represented by recursion
relations\index{recursive dynamics} \cite{pc-SB01,pc-SBP02} in discrete time steps. Let $U_{t}$ be the fraction of
fibers in the initial bundle that survive after time step $t$, where
time step indicates the number of occurrence of stress redistribution.
Then the redistributed stress after $t$ time step becomes \begin{equation}
\sigma_{t}=\frac{\sigma}{U_{t}};\label{pc-2}\end{equation}

\noindent and after $t+1$ time steps the surviving fraction of fiber
is \begin{equation}
U_{t+1}=1-P(\sigma_{t});\label{pc-3}\end{equation}

\noindent where $P(\sigma_{t})$ is the cumulative probability of
corresponding density distribution $p(x)$: $P(\sigma_{t})=\int_{0}^{\sigma_{t}}p(x)dx.$
Now using Eq. (\ref{pc-2}) and Eq. (\ref{pc-3}) we can write the
recursion relations which show how $\sigma_{t}$ and $U_{t}$ evolve
in discrete time: \begin{equation}
\sigma_{t+1}=\frac{\sigma}{1-P(\sigma_{t})};\sigma_{0}=\sigma\label{pc-4}\end{equation}
and \begin{equation}
U_{t+1}=1-P(\sigma/U_{t});U_{0}=1.\label{pc-5}\end{equation}

At the equilibrium or steady state $U_{t+1}=U_{t}\equiv U^{*}$ and
$\sigma_{t+1}=\sigma_{t}\equiv\sigma^{*}$. This is a fixed point\index{fixed-point}
of the recursive dynamics. Eq. (\ref{pc-4}) and Eq. (\ref{pc-5})
can be solved at the fixed point for some particular distribution.
 At the fixed-point,
the solutions \cite{pc-SB01,pc-SBP02,pc-PSB03} assume universal form \begin{equation}
\sigma^{*}(\sigma)=C-(\sigma_{c}-\sigma)^{1/2};\label{pc-sigma}\end{equation}

\begin{equation}
U^{*}(\sigma)=C+(\sigma_{c}-\sigma)^{1/2};\label{pc-u}\end{equation}

\noindent {where $\sigma_{c}$ is the critical stress and $C$ is a constant.
From the recursions and their solutions we can derive the following
response quantities:}

\textbf{Susceptibility}{: Which
is defined as the amount of change\index{susceptibility} in $U$ when the external stress
changes by a infinitesimal amount and can be derived as \begin{equation}
\chi=\left|\frac{dU^{*}(\sigma)}{d\sigma}\right|=\frac{1}{2}(\sigma_{c}-\sigma)^{-\beta};\beta=\frac{1}{2}.\label{pc-ky}\end{equation}
 }

\textbf{Relaxation time}{: This
is the number of step\index{relaxation} the bundle needs to come to a stable state after
the application of an external stress. From the solution of recursive
dynamics we get \cite{pc-SB01,pc-SBP02} near $\sigma_{c}$\begin{equation}
\tau\propto(\sigma_{c}-\sigma)^{-\theta};\theta=\frac{1}{2}.\label{pc-tau}\end{equation}
 }

\textbf{Inclusive avalanche}{:
This is the amount of avalanche per step of stress redistribution\begin{equation}
\frac{dU}{dt}=U_{t}-U_{t+1}\label{pc-inclu}\end{equation}
}

\noindent{At $\sigma=\sigma_{c}$, the dynamics
becomes {}``critically slow'' as \begin{equation}
U_{t}\sim t^{-\gamma};\gamma=1;\label{pc-crit}\end{equation}
}

\noindent{which suggests that at $\sigma_{c}$ inclusive
avalanches\index{inclusive avalanche} follow a power law (exponent $-2$) decay with time.}

\subsubsection{Prediction of global failure point}

{\large (A)} \textbf{{\large Using
$\chi$ and $\tau$}}{\large \par}

\begin{center}{\includegraphics[%
  width=1.8in,
  height=1.6in]{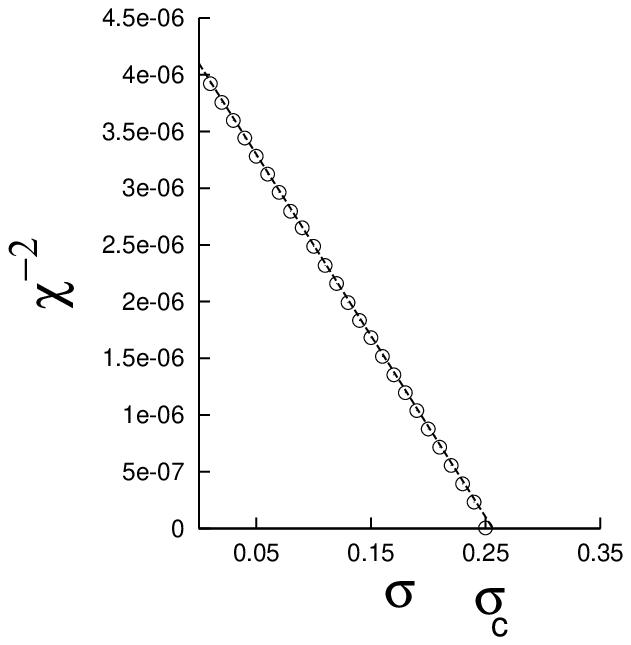}\hskip.5in\includegraphics[%
  width=1.8in,
  height=1.6in]{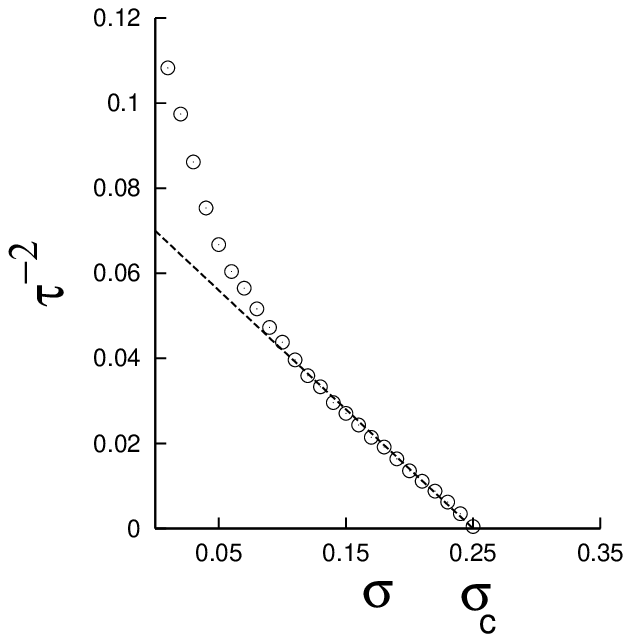}}\end{center}

{\small Fig. 2: Variation of $\chi^{-2}$ and $\tau^{-2}$
with applied stress for a bundle having $N=50000$ fibers and averaging
over $1000$ samples. We consider uniform distribution of fiber threshold. }{\small \par}
\vskip.1in
We found that susceptibility ($\chi$) and relaxation time ($\tau$)
follow power laws with external stress and both of them diverge at
the critical stress. Therefore if we plot $\chi^{-2}$ and $\tau^{-2}$
with external stress, we expect a linear fit near critical point\index{critical point} and
the straight lines should touch $X$ axis at the critical stress.
We indeed found similar behavior (Fig. 2) in simulation experiments. 

For application, it is always important that such prediction can be
done in a single sample. We have performed the simulation taking a
single bundle having very large number of fibers and we observe similar
response of $\chi$ and $\tau$. The prediction\index{prediction} of failure point is
also quite satisfactory (Fig. 3).

\begin{center}\includegraphics[%
  width=1.8in,
  height=1.6in]{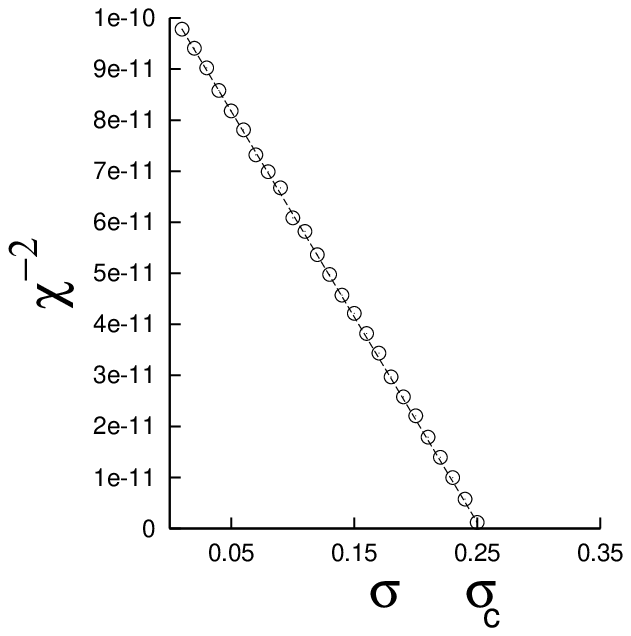}\hskip.5in\includegraphics[%
  width=1.8in,
  height=1.6in]{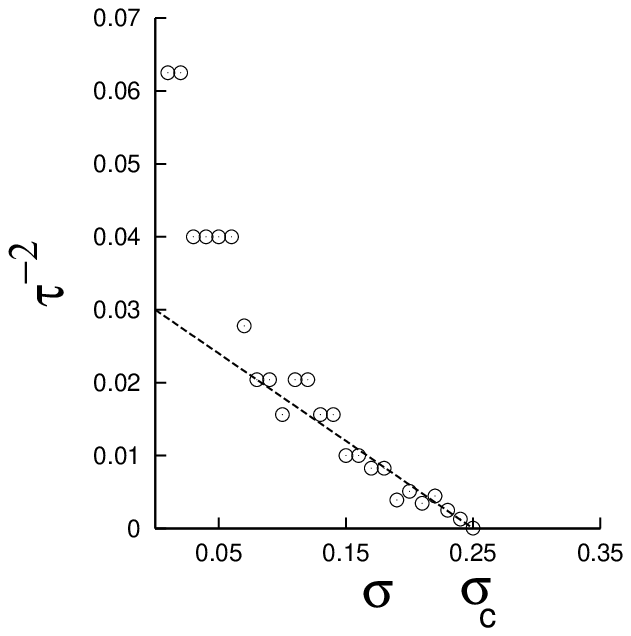}\end{center}

{\small Fig. 3: Variation of $\chi^{-2}$ and $\tau^{-2}$
with applied stress for a single bundle having $N=10000000$ fibers
with uniform distribution of fiber threshold. }{\small \par}

\vskip.1in

{\large (B)} \textbf{{\large Using
inclusive avalanche}}{\large \par}
\vskip.1in

\begin{center}{\includegraphics[%
  width=1.8in,
  height=1.6in]{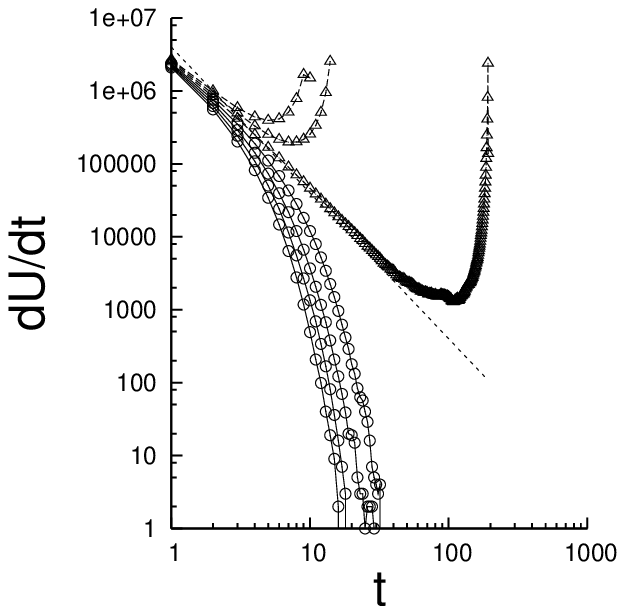}\hskip.5in\includegraphics[%
  width=1.8in,
  height=1.6in]{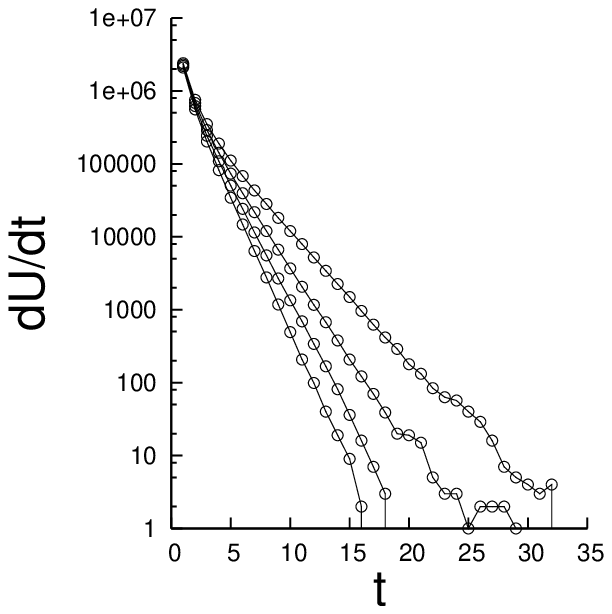}}\end{center}

{\small Fig. 4: Log-log plot of inclusive avalanche
with step of load redistribution for $7$ different stress values [left] and 
log-normal plot of the same for $4$ different stress values below the critical 
stress [right]. The simulation has been performed for a single bundle with 
$N=10000000$ fibers having uniform distribution of fiber threshold. }{\small \par}
\vskip.1in
When we put a big load on a material body, sometimes it becomes important
to know whether that body can support the load or not. The similar
question can be asked in FBM. We found that if we record the inclusive
avalanche, i.e, the amount of failure in each load redistribution
-then the pattern of inclusive burst clearly shows whether the bundle
is going to fail or not. For any stress below the critical state,
inclusive avalanche follow exponential decay (Fig. 4 (right)) with time 
step and for
stress values above critical stress it is a power law followed by
a gradual rise (Fig. 4 (left)). Clearly at critical stress it follows 
a robust power law with exponent $-2$ that we already get analytically. 

\vskip.1in
{\large (C)} \textbf{{\large Using
avalanche distribution}}{\large \par}

\begin{center}\includegraphics[%
  width=2.2in,
  height=1.8in]{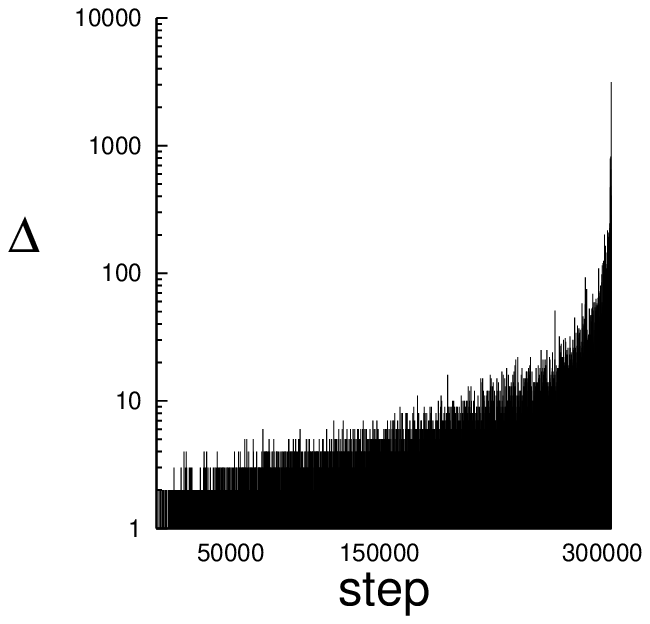}\end{center}

{\small Fig. 5: Magnitude of avalanches with step
of load increment in FBM. }{\small \par}

\vskip.1in

An {}``Avalanche'' is the amount of failure\index{avalanche} occurs as the system
moves from one stable state to the next stable state when the load
is increased quasi-statically. We can simply measure it by counting
the failure elements between two consecutive load increment. Such
a series of avalanches has been shown in Fig. 5 up to the final failure
point. If we record all the avalanches till final failure, the avalanche
distribution follows a universal power law\index{power-law} \cite{pc-HH92}. This avalanches
can be recorded experimentally measuring the acoustic emissions during
fracture-failure of materials. We want to study whether the avalanche
distribution changes if we start gathering the avalanches from some
intermediate states of the breaking process (Fig. 6 (left)). In simulations
we see that the exponent of avalanche distribution shows a crossover\index{crossover}
between two values ($-5/2$ ans $-3/2$) and the crossover point (length) 
depends on the starting position ($x_{0}$) of our measurement (Fig. 6(right)). 

\noindent \begin{center}{\includegraphics[%
  width=1.8in,
  height=1.6in]{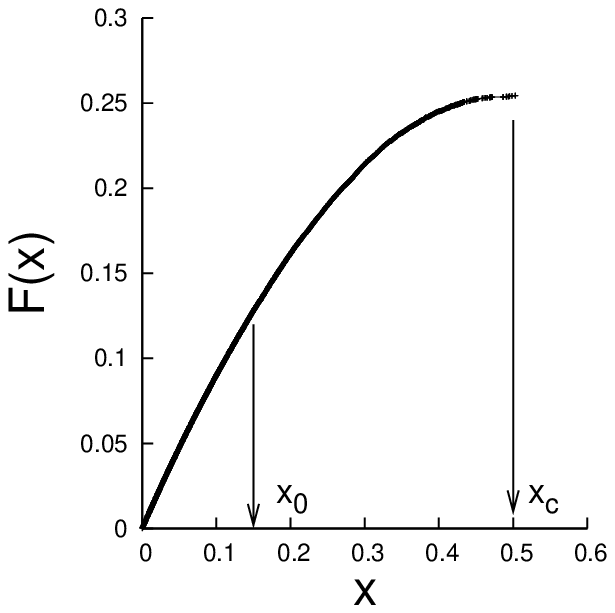}\hskip.5in\includegraphics[%
  width=1.8in,
  height=1.6in]{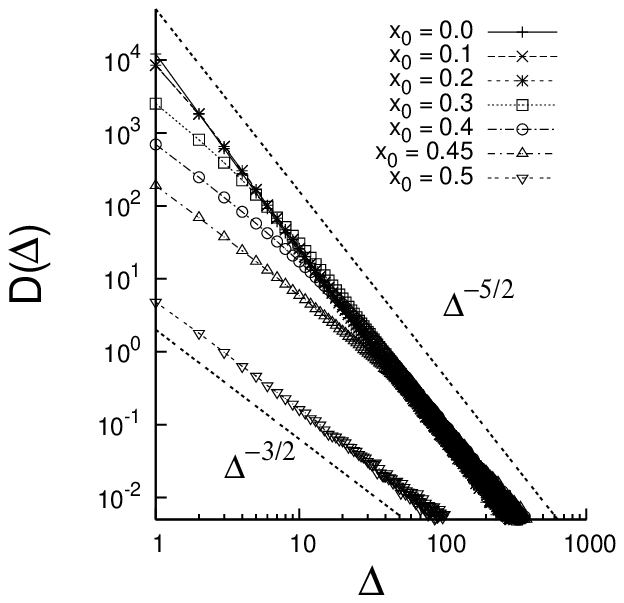}}\end{center}

{\small Fig. 6: Average force versus elongation curve
(left) and simulation results of avalanche distributions for different
starting position (right) in FBM. }{\small \par}

\vskip.1in

Now we are going to explain the above observation analytically for
quasi-static loading. For a bundle of many fibers the average number
of bursts of magnitude $\Delta$ is given by \cite{pc-HH92}\begin{equation}
\frac{D(\Delta)}{N}=\frac{\Delta^{\Delta-1}}{\Delta!}\int_{0}^{x_{c}}p(x)\left[1-xp(x)/Q(x)\right]\left[xp(x)/Q(x)\right]^{\Delta-1}\end{equation}

\begin{equation}\qquad\times\exp\left[-\Delta xp(x)/Q(x)\right]dx,\label{pc-6}\end{equation}

\noindent where $Q(x)=\int_{x}^{\infty}p(y)\; dy$ is the fraction
of total fibers with strength exceeding $x$ and $x_{c}$ is the critical value 
beyond which the bundle fails instantly. For uniform 
distribution\index{uniform distribution} (having upper bound $x_{m}$), if we 
start recording avalanches from a intermediate point $x_{0}$, the avalanche 
distribution becomes 
\begin{equation}
\frac{D(\Delta)}{N}=\frac{\Delta^{\Delta-1}}{\Delta!(x_{m}-x_{0})}\times\int_{x_{0}}^{x_{c}}\frac{x_{m}-2x}{x}\left[\frac{x}{x_{m}-x}\, e^{-x/(x_{m}-x)}\right]^{\Delta}\; dx.\label{pc-7}\end{equation}

Introducing the parameter $\epsilon=\frac{x_{c}-x_{0}}{x_{m}}$ and
a new integration variable $z=\frac{x_{m}-2x}{\epsilon(x_{m}-x)},$
we obtain \begin{equation}
\frac{D(\Delta)}{N}=\frac{2\Delta^{\Delta-1}e^{-\Delta}\,\epsilon^{2}}{\Delta!(1+2\epsilon)}\times\int_{0}^{4/(1+2\epsilon)}\frac{z}{(1-\epsilon z)(2-\epsilon z)^{2}}\; e^{\Delta[\epsilon z+\ln(1-\epsilon z)]}\; dz.\label{pc-8}\end{equation}

\begin{flushleft}For small $\epsilon$, i.e., close to the critical
threshold distribution, we can expand \begin{equation}
\epsilon z+\ln(1-\epsilon z)=-{\textstyle \frac{1}{2}}\epsilon^{2}z^{2}-{\textstyle \frac{1}{3}}\epsilon^{3}z^{3}+\ldots,\label{pc-9}\end{equation}
 with the result\end{flushleft}

\begin{equation}
\frac{D(\Delta)}{N}\simeq\frac{\Delta^{\Delta-1}e^{-\Delta}\epsilon^{2}}{2\Delta!}\times\int_{0}^{4}e^{-\Delta\epsilon^{2}z^{2}/2}\; zdz=\frac{\Delta^{\Delta-2}e^{-\Delta}}{2\Delta!}\left(1-e^{-8\epsilon^{2}\Delta}\right).\label{pc-10}\end{equation}
 Using Stirling approximation $\Delta!\simeq\Delta^{\Delta}e^{-\Delta}\sqrt{2\pi\Delta},$
this becomes

\begin{flushleft}\begin{equation}
\frac{D(\Delta}{N}\simeq(8\pi)^{-1/2}\Delta^{-5/2}\left(1-e^{-\Delta/\Delta_{c}}\right),\label{pc-11}\end{equation}
 with \begin{equation}
\Delta_{c}=\frac{1}{8\epsilon^{2}}=\frac{x_{m}^{2}}{8(x_{c}-x_{0})^{2}}.\label{pc-12}\end{equation}
 Clearly, there is a crossover \cite{pc-PHH-05} at a burst length around
$\Delta_{c}$, so that \begin{equation}
\frac{D(\Delta)}{N}\simeq\left\{ \begin{array}{cl}
(8/\pi)^{1/2}\epsilon^{2}\;\Delta^{-3/2} & \mbox{ for }\Delta\ll\Delta_{c}\\
(8\pi)^{-1/2}\Delta^{-5/2} & \mbox{ for }\Delta\gg\Delta_{c}\end{array}\right.\label{pc-13}\end{equation}
 For $x_{0}=0.40x_{m}$, we have $\Delta_{c}=12.5$ uniform distribution.
We found a clear crossover near $\Delta=\Delta_{c}=12.5$ in the simulation
experiment (Fig. 7).\end{flushleft}

\noindent \begin{center}{\includegraphics[%
  width=2.0in,
  height=1.8in]{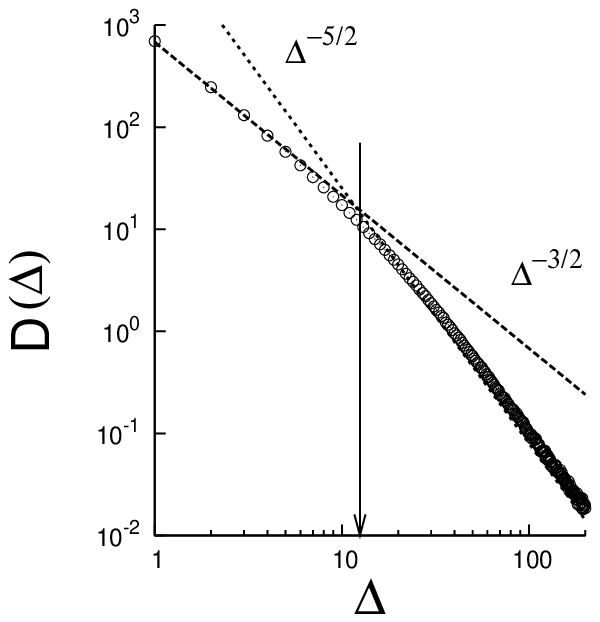}}\hskip.5in\includegraphics[%
  width=2.0in,
  height=1.8in]{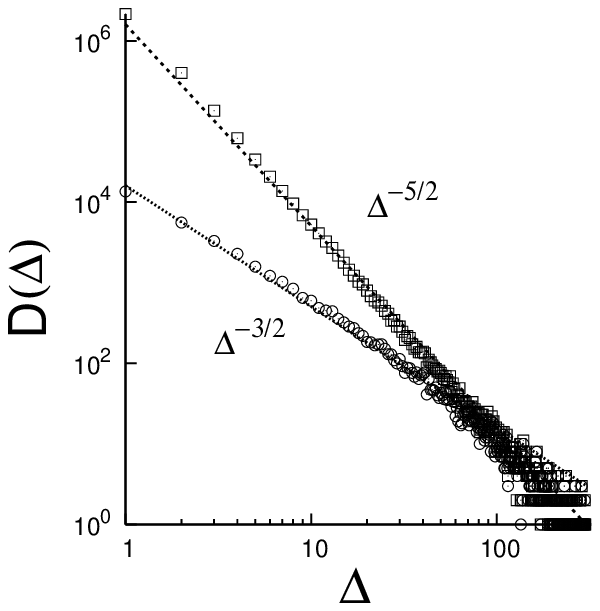}\end{center}

 {\small Fig. 7: Crossover
in avalanche distributions: bundle with $10^{6}$ fiber and
$x_{0}=0.8x_{c}$ (averaging over $20000$ samples) [left] and
 a single bundle with $10^{7}$ fiber; $x_{0}=0$
 (squares) and  $x_{0}=0.9x_{c}$
(circles) [right]. Dotted straight lines are the
best fits to the power laws. }{\small \par}

\vskip.1in

The simulation results shown in the figures are based on \textit{averaging}
over a large number of samples. For applications
it is important that crossover signals are seen also in a single sample.
We show (Fig. 7) that equally clear power laws are seen in a \textit{single}
fiber bundle when $N$ is large. 

This crossover phenomenon is not limited to the uniform threshold
distribution. The $\xi=3/2$ power law \cite{pc-PHH-05} in the burst
size distribution dominates over the $\xi=5/2$ power law whenever
a threshold distribution is non-critical, but close to criticality.
Therefore, the magnitude of the crossover length correctly inform us how
far the system is from the global failure point.

\subsection{\noindent{Electrical networks within a voltage
difference: Random fuse model}}

\vskip.2in

\begin{center}{\includegraphics[%
  width=2.0in,
  height=1.8in]{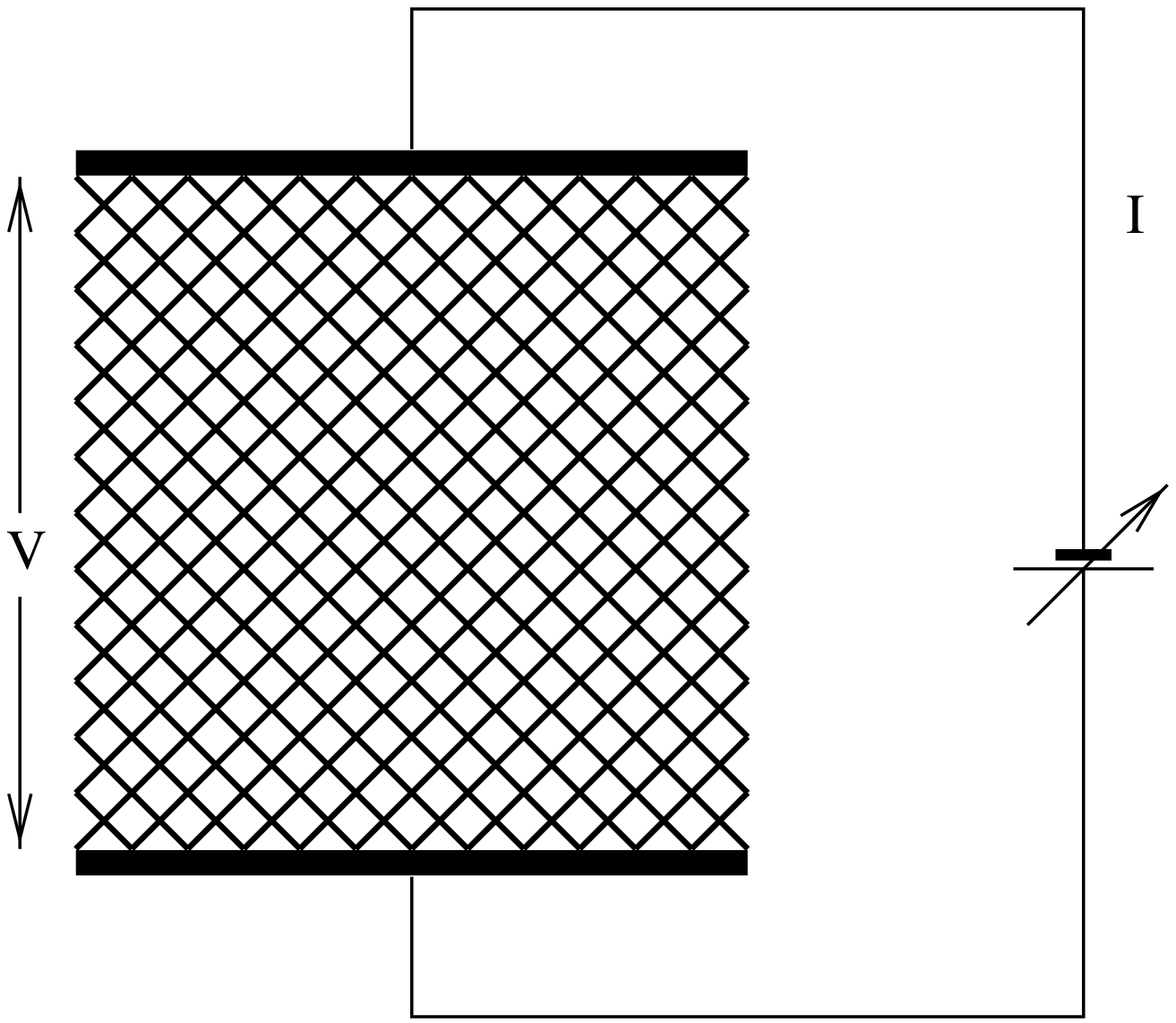}\hskip.5in\includegraphics[%
  width=2.0in,
  height=1.8in]{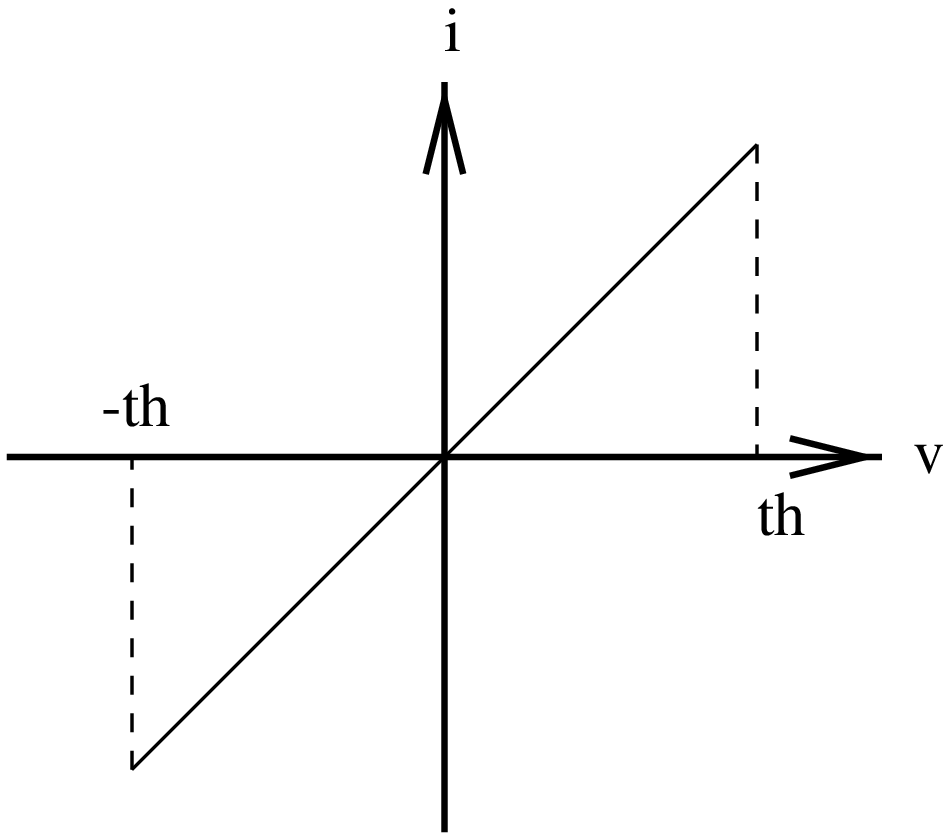}}\end{center}

{\small Fig. 8: Random fuse model (left) and the current-response
of a single Ohmic resistor (right). }{\small \par}

\vskip.1in

Random fuse model\index{fuse model} \cite{pc-books} describes the breakdown phenomena
in electrical networks. It consists of a lattice in which each bond
is a fuse, i.e., an Ohmic resistor as long as the electric current
it carries is below a threshold value. If the threshold is passed,
the fuse burns out irreversibly (Fig. 8 (right)). The threshold ($th$) of 
each bond is drawn from an uncorrelated distribution $p(th)$. 
All fuses have the same resistance.
The lattice is a two-dimensional square one placed at 45$^{\circ}$
with regards to the bus bars (Fig. 8 (left)) and an increasing current 
is passed through it. Numerically, the Kirchhoff equations are solved at 
each point of the system.

\begin{center}{\includegraphics[%
  width=1.8in,
  height=1.6in]{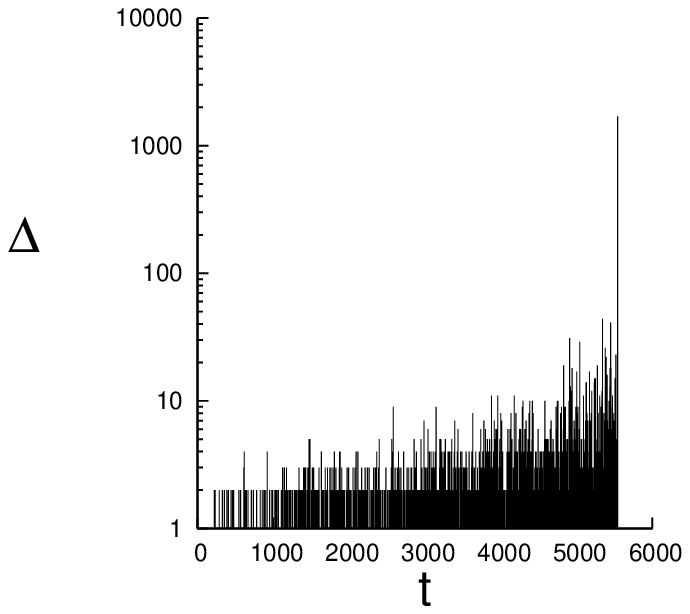}\hskip.5in\includegraphics[%
  width=1.8in,
  height=1.6in]{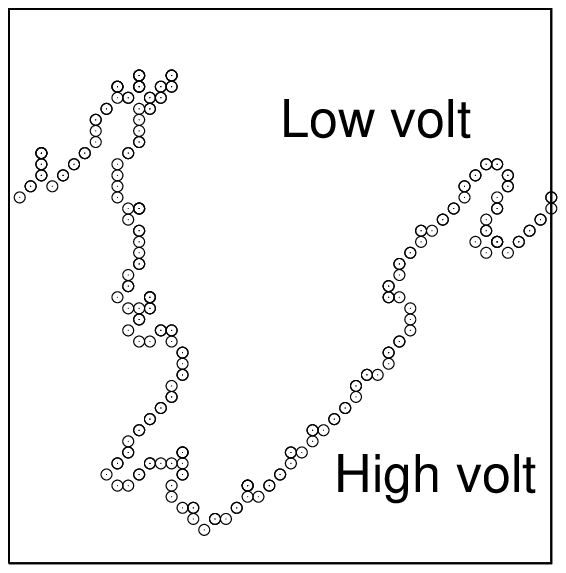}}\end{center}

{\small Fig. 9: Avalanches in fuse model (left) and the
ultimate fractured state (right).}{\small \par}

\vskip.1in

For the entire failure process avalanches\index{avalanche} {of different
sizes occur (Fig. 9 (left)) and the system finally comes to a point
where no current passes through the system (Fig. 9 (right))- we call
it the global failure point. The average avalanche size has been measured
\cite{pc-Phase T} for such system and it follows a power law with increasing
current or voltage:}

\begin{center}{$m\sim(I_{c}-I)^{-\gamma}$ or $m\sim(V_{c}-V)^{-\gamma}$;$\gamma=1/2$}\end{center}

Therefore if we plot $m^{-2}$ with $I$ or $V$ we can expect a linear
fit which touches $X$ axis at the critical current value $I_{c}$
or critical voltage value $V_{c}$ (Fig. 10). 

\begin{center}{\includegraphics[%
  width=1.8in,
  height=1.6in]{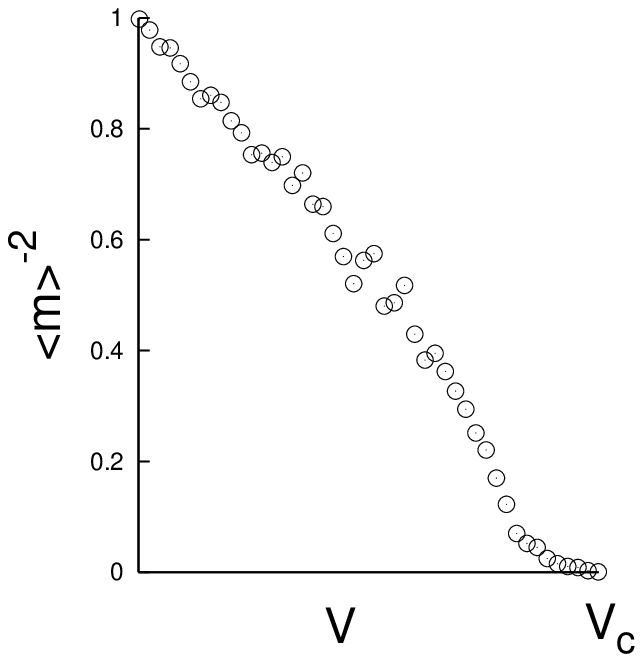}\hskip.5in\includegraphics[%
  width=1.8in,
  height=1.6in]{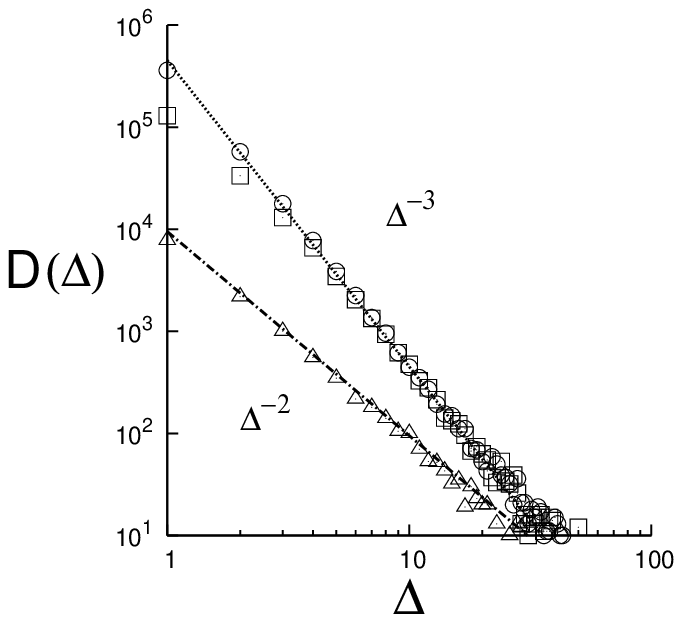}}\end{center}

{\small Fig. 10: Plot of} $m^{-2}$ with $V$ {\small (left) and
the crossover in avalanche distributions in a fuse model (right).}{\small \par}

\vskip.1in

If we record all the avalanches, the avalanche distribution follows
a universal power law with $\xi\simeq3$. But what will happen if
we start recording the burst at some intermediate state of the failure
process? We observe \cite{pc-PHH-05} that for a system of size $100\times100$,
$2097$ fuses blow on the average before catastrophic failure sets
in. When measuring the burst distribution only after the first $2090$
fuses have blown, a different power law is found\index{crossover}, this time with $\xi=2$.
After $1000$ blown fuses, on the other hand, $\xi$ remains the same
as for the histogram recording the entire failure process (Fig. 10).

\subsection{\noindent {SOC models of sandpile}}

\vskip.4in

\begin{center}{\includegraphics[%
  width=2.5in,
  height=2in]{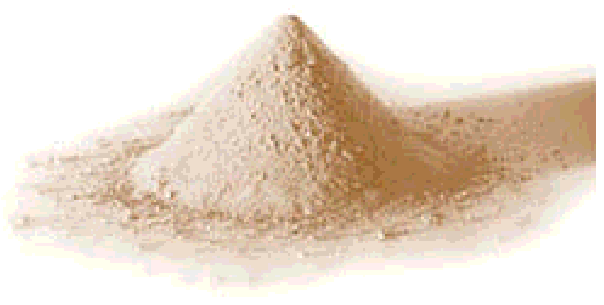}\vskip.2in\includegraphics[%
  width=4.5in,
  height=1.5in]{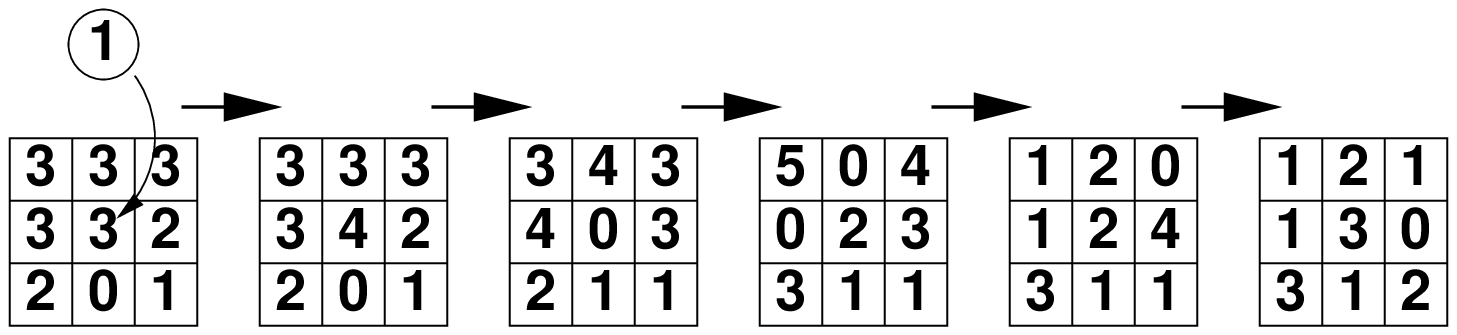}}\end{center}

{\small Fig. 11: A real sandpile and the sandpile
model \cite{pc-Manna} on a square lattice.}{\small \par}

\vskip.1in

Growth of a natural sandpile\index{sandpile} is a nice example of self-organised criticality\index{SOC} [15-19].
If sand grains are added continuously on a small pile, the system
gradually approaches towards a state at the boundary between stable
and unstable states where the system shows long-range spatio-temporal
fluctuations similar to those observed in equilibrium critical phenomena.
This special state has been identified as the critical state\index{critical point} of the
pile where the response to addition of sand grains becomes unpredictable:
Avalanche of any size is equally probable at this state. Therefore, 
 we can expect system spanning avalanche (global failure) only at this 
critical state.

\subsubsection{\noindent{BTW model and Manna model}}

Bak et. al. \cite{pc-BTW} proposed a sandpile model\index{BTW model} on square lattice
which captures correctly the properties of a natural sandpile. At
each lattice site $(i,j)$, there is an integer variable $h_{i,j}$
which represents the height of the sand column at that site. A unit
of height (one sand grain) is added at a randomly chosen site at each
time step and the system evolves in discrete time. The dynamics starts
as soon as any site $(i,j)$ has got a height equal to the threshold
value ($h_{th}$= $4$): that site topples, i.e., $h_{i,j}$ becomes
zero there, and the heights of the four neighbouring sites increase
by one unit \begin{equation}
h_{i,j}\rightarrow h_{i,j}-4,h_{i\pm1,j}\rightarrow h_{i\pm1,j}+1,h_{i,j\pm1}\rightarrow h_{i,j\pm1}+1.\label{pc-14}\end{equation}
 If, due to this toppling at site $(i,j)$, any neighbouring site
become unstable (its height reaches the threshold value), the
 same dynamics follows. Thus the process continues till all sites
become stable ($h_{i,j}<$ $h_{th}$ for all $(i,j)$). When toppling
occurs at the boundary of the lattice, extra heights get off the lattice
and are removed from the system. With continuous addition of unit
height (sand grain) at random sites of the lattice, the avalanches
(toppling) get correlated over longer and longer ranges and the average
height ($h_{av}$) of the system grows with time. Gradually the correlation
length ($\xi$) becomes of the order the system size $L$ as the
system attains the critical average height $h_{c}(L)$. On
average, the additional height units start leaving the system and
the average height remains stable there (see Fig. 12 (a)). The distributions
of the avalanche sizes and the corresponding life times follow robust
power laws \cite{pc-Dhar}, hence the system becomes critical here. 

We can perform a finite size scaling fit $h_{c}(L)=h_{c}(\infty)+{\textrm{C}}L^{-1/\nu}$
(by setting $\xi$ $\sim$ $\mid h_{c}(L)-h_{c}(\infty)\mid^{-\nu}=L$),
where C is a constant, with $\nu\simeq1.0$ gives $h_{c}\equiv h_{c}(\infty)\simeq2.124$
(see inset of Fig. 12 (a)). Similar finite size scaling fit with $\nu=1.0$
gave $h_{c}(\infty)\simeq2.124$ in earlier large scale simulations
\cite{pc-Manna}. 

\begin{center}{\includegraphics[%
  width=2.5in,
  height=2.0in,
  angle=-90]{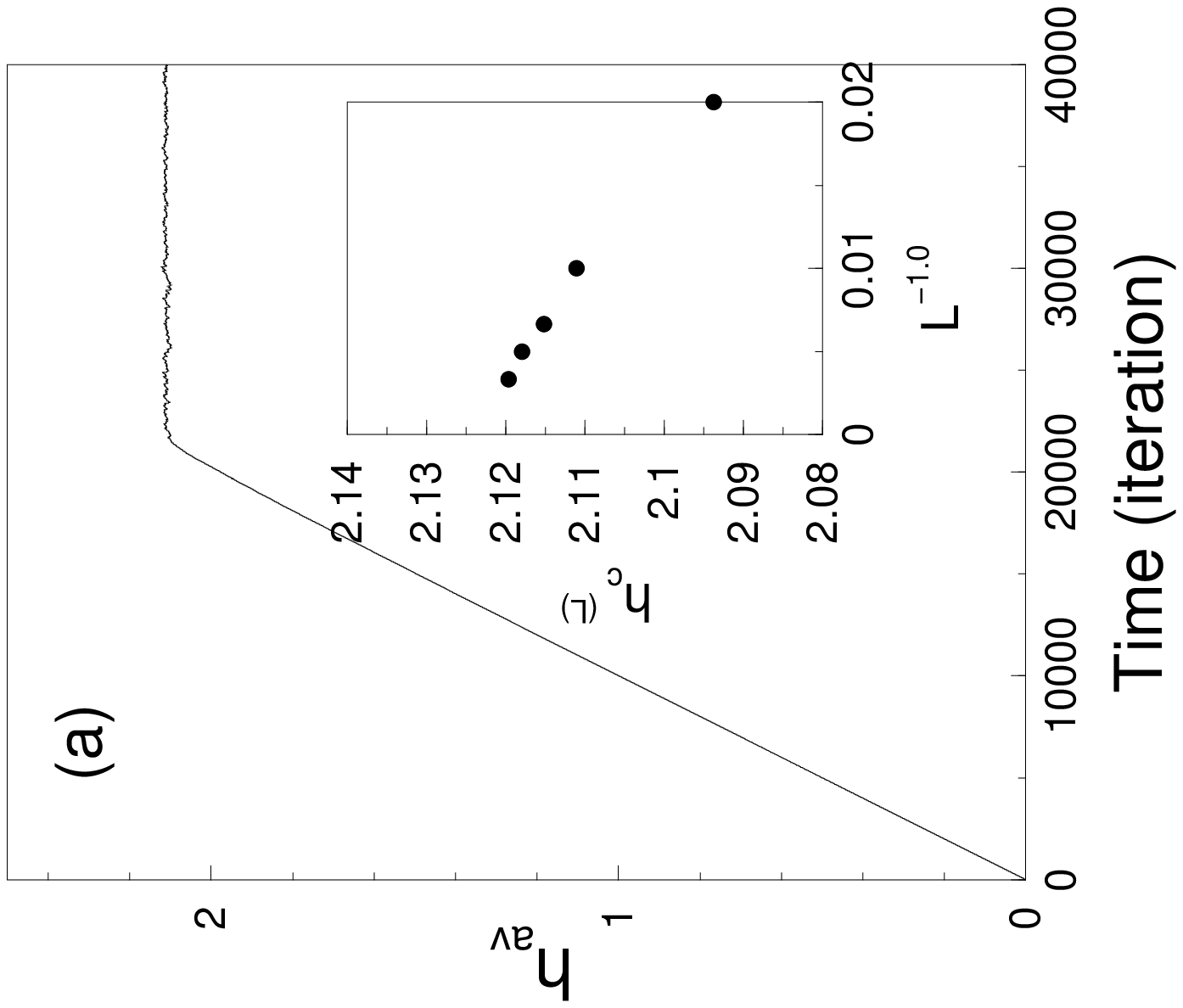}\hskip.2in \includegraphics[%
  width=2.5in,
  height=2.0in,
  angle=-90]{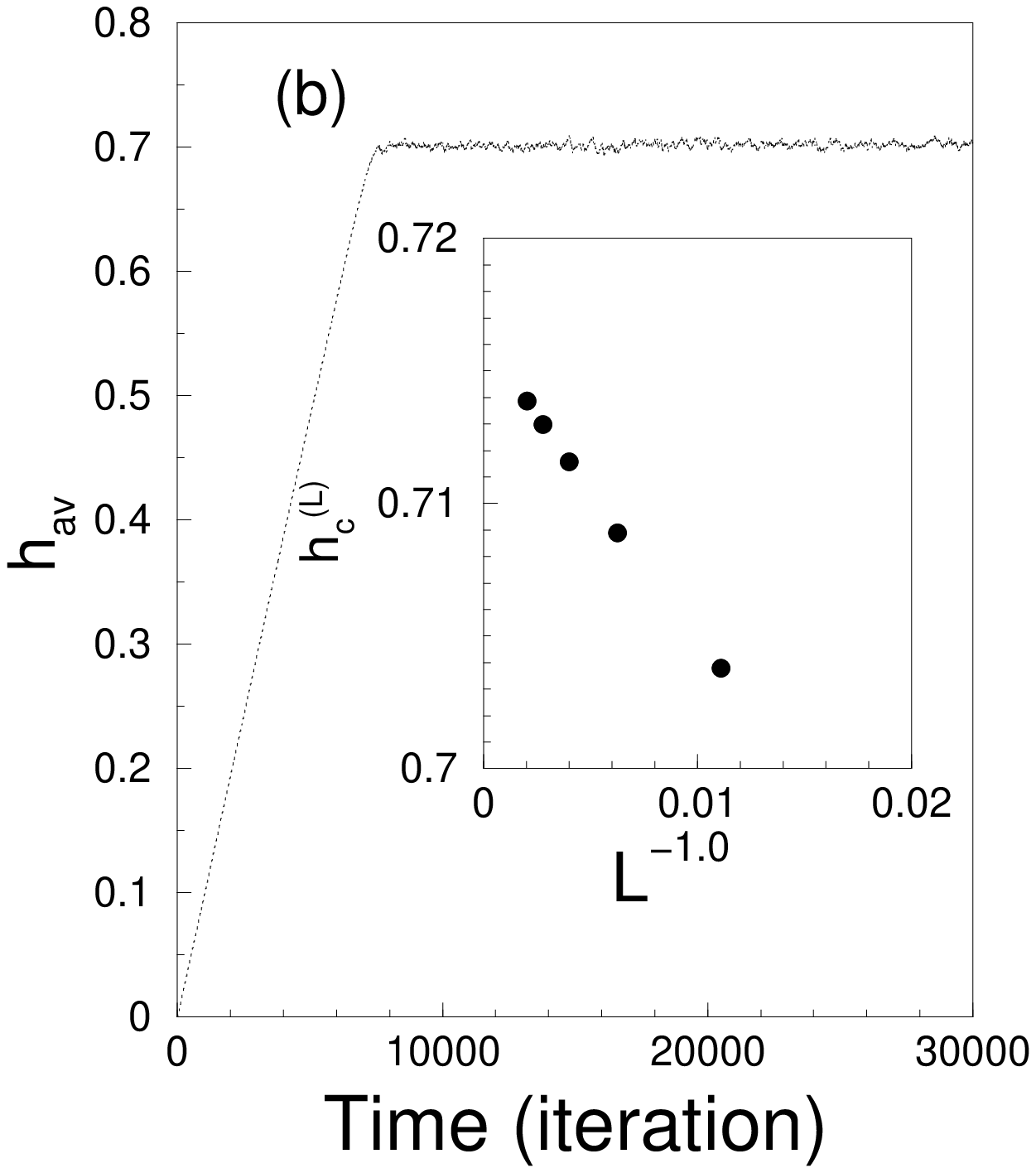}}\end{center}

{\small Fig. 12: Growth of BTW (a) and Manna (b) sandpiles.
Inset shows the system size dependence of the critical height.}{\small \par}

\vskip.1in

Manna proposed the stochastic sand-pile model\index{Manna model} \cite{pc-Manna} by introducing
randomness in the dynamics of sand-pile growth in two-dimensions.
Here, the critical height is $2$. Therefore at each toppling, two
rejected grains choose their host among the four available neighbours
randomly with equal probability. After constant adding of sand grains,
the system ultimately settles at a critical state having height $h_{c}$.
A similar finite size scaling fit $h_{c}(L)=h_{c}(\infty)+{\textrm{C}}L^{-1/\nu}$
gives $\nu\simeq1.0$ and $h_{c}\equiv h_{c}(\infty)\simeq0.716$
(see inset of Fig. 12 (b)). This is close to an earlier estimate $h_{c}\simeq0.71695$
\cite{pc-Ves}, made in a somewhat different version of the model. The
avalanche size distribution has got power laws similar to the BTW
model, however the exponent seems to be different \cite{pc-Manna,pc-Dhar},
compared to that of BTW model.

\subsubsection{{Sub-critical response: Precursors }}

We are going to investigate the behavior of BTW and Manna sandpiles when thay 
are away from the critical state ($h<h_{c}$).
At an average height $h_{av}$, when all sites of the system have
become stable\index{sub-critical} (dynamics have stopped), a fixed number of height units
$h_{p}$ (pulse of sand grains) is added at any central point of the
system \cite{pc-SB01,pc-AC96}. Just after this addition, the local dynamics starts and it
takes a finite time or iterations to return back to the stable state
($h_{i,j}<h_{th}$ for all $(i,j)$) after several toppling events.
{We measure the response parameters: $\Delta$ $\rightarrow$
number of toppling $\tau$ $\rightarrow$ number of iteration and
$\xi$ $\rightarrow$ correlation length} which is the distance of
the furthest toppled site from the site where $h_{p}$ has been dropped.

\newpage

{(A)} \textbf{{In BTW model}}

\vskip.2in

{We choose $h_{p}=4$ for BTW model to ensure toppling
at the target site. We found that all the response parameters follow
power law as $h_{c}$ is approached:$\Delta\propto(h_{c}-h_{av})$$^{-\lambda}$,
$\tau\propto(h_{c}-h_{av})^{-\mu}$, $\xi\propto(h_{c}-h_{av})^{-\nu}$;
$\lambda\cong2.0$, $\mu\cong1.2$ and $\nu\cong1.0$. Now if we plot
$\Delta^{-1/\lambda}$, $\tau^{-1/\mu}$ and $\xi^{-1/\nu}$ with
$h_{av}$ all the curve follow straight line and they should touch
the x axis at $h_{av}=h_{c}$. Therefore by a proper extrapolation
we can estimate the value of $h_{c}$ and we found $h_{c}$ $=2.13\pm.01$
 (Fig. 13) which agree well with direct estimates of the same.}

\begin{center}{\includegraphics[%
  width=2.5in,
  height=2.0in,
  angle=-90]{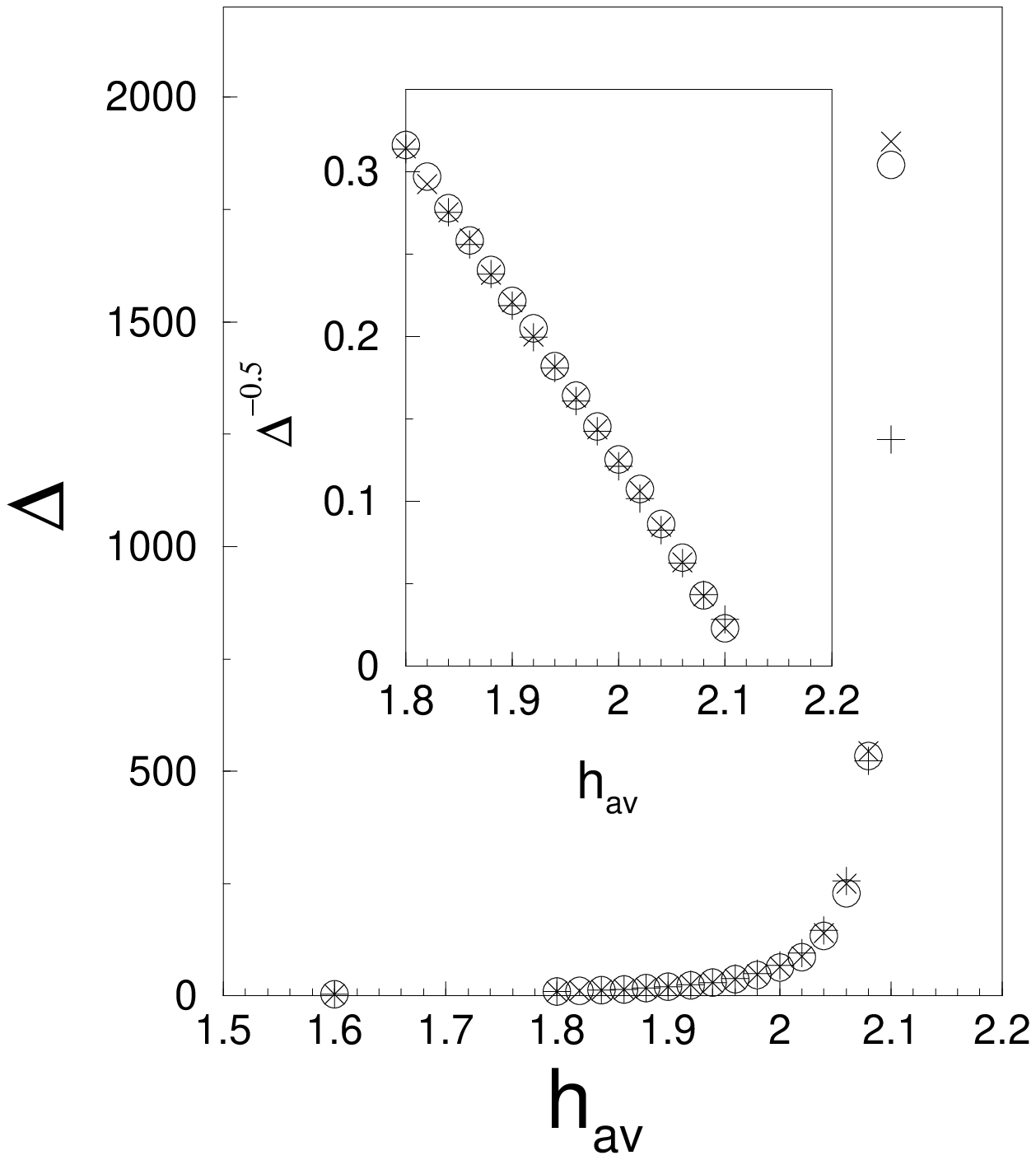}\hskip.2in\includegraphics[%
  width=2.5in,
  height=2.0in,
  angle=-90]{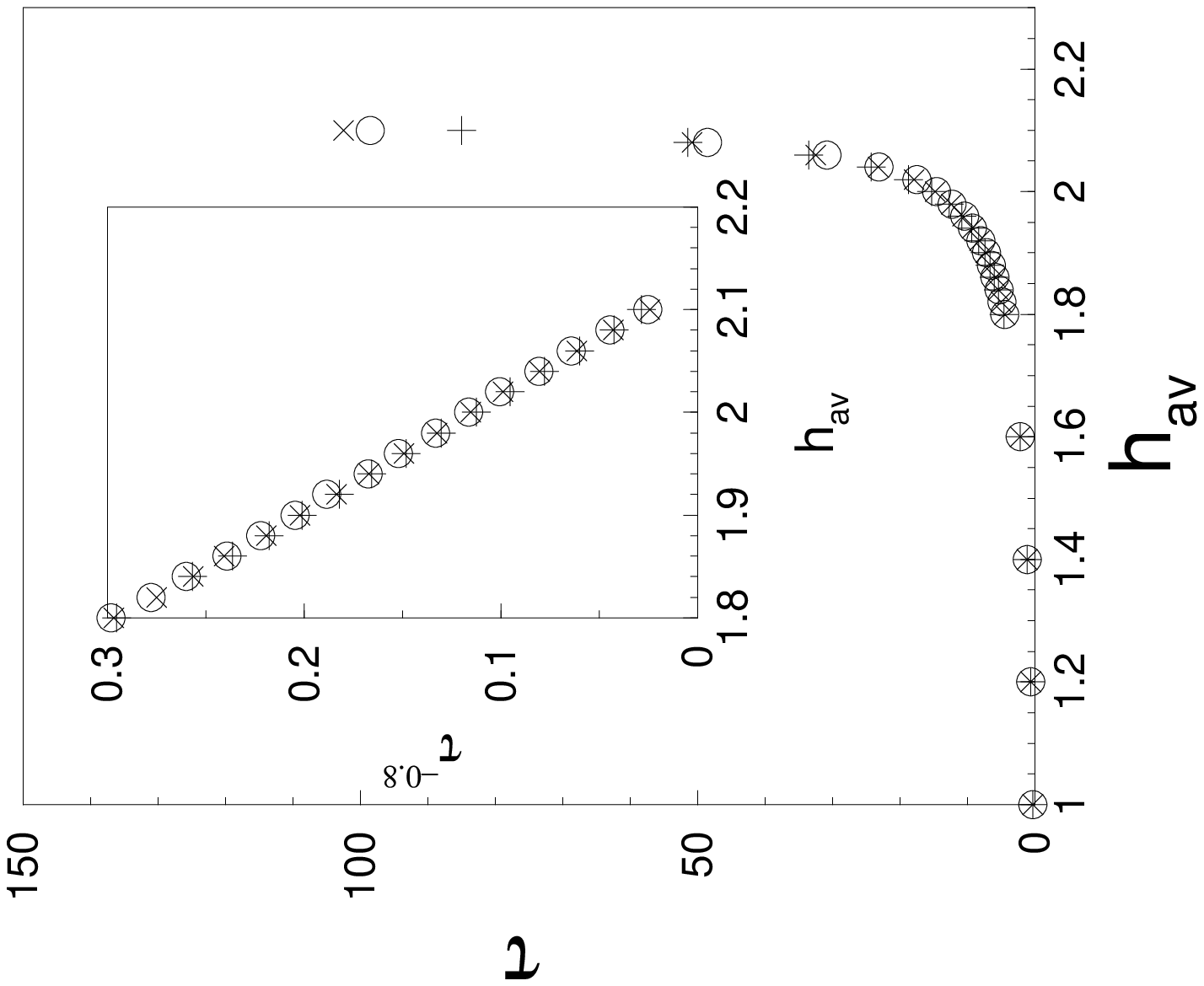}}\end{center}

\begin{center}{\includegraphics[%
  width=2.5in,
  height=2.5in,
  angle=-90]{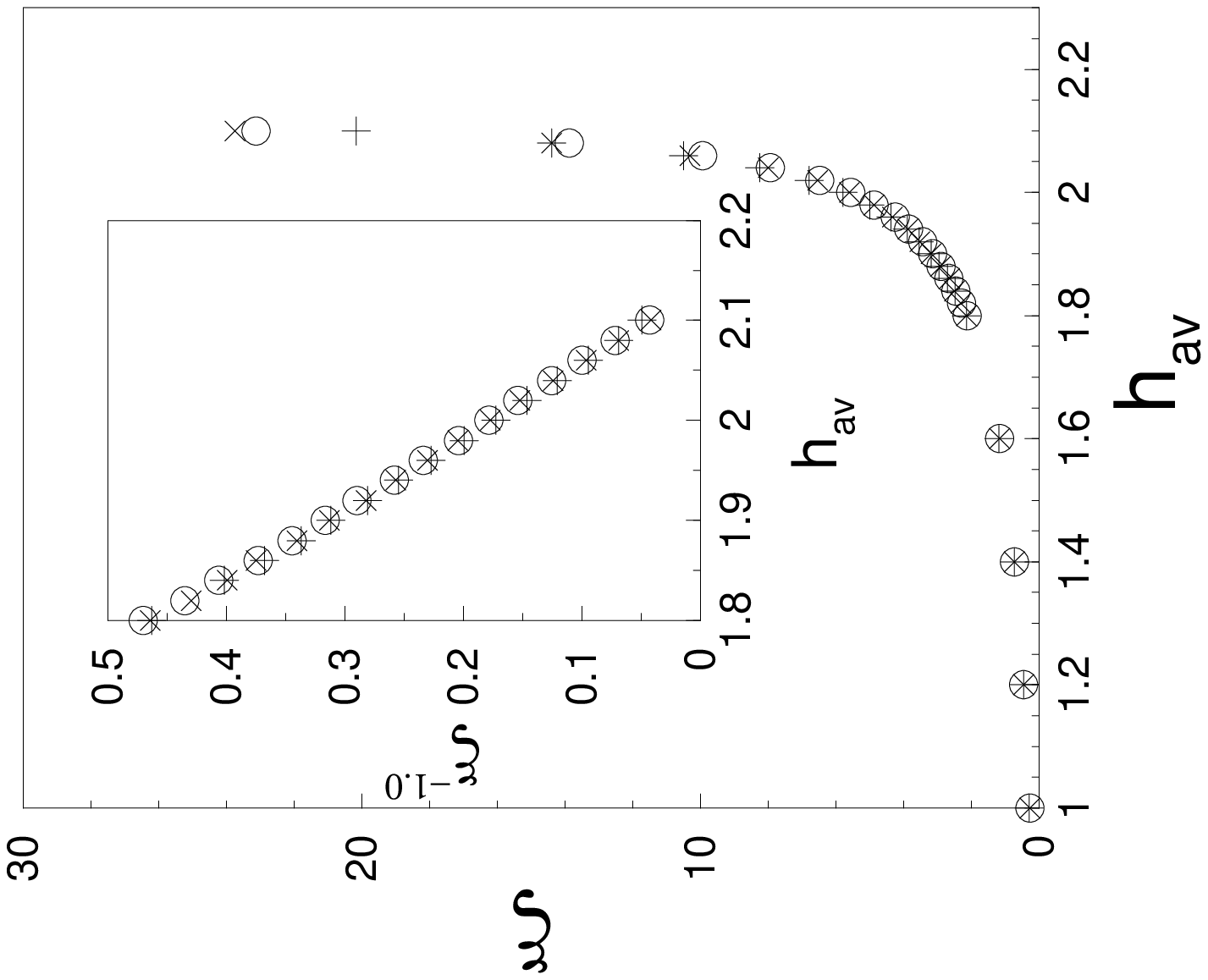}}\end{center}

{\small Fig. 13: Precursor parameters and prediction
of critical point in BTW model.}{\small \par}

\newpage

{(B)} \textbf{{In Manna model}}

\vskip.2in

{Obviously we have to choose $h_{p}=2$ for Manna
model to ensure toppling at the target site. Then we measured all
the response parameters and they seem to follow power law as $h_{c}$
is approached:$\Delta\propto(h_{c}-h_{av})$$^{-\lambda}$, $\tau\propto(h_{c}-h_{av})^{-\mu}$,
$\xi\propto(h_{c}-h_{av})^{-\nu}$; $\lambda\cong2.0$, $\mu\cong1.2$
and $\nu\cong1.0$. As in BTW model, all the curve follow straight
line if we plot $\Delta^{-1/\lambda}$, $\tau^{-1/\mu}$ and $\xi^{-1/\nu}$
with $h_{av}$ and proper extrapolations estimate the value of $h_{c}$
$=0.72\pm.01$ which is again a good estimate (Fig. 14).}

\begin{center}{\includegraphics[%
  width=2.5in,
  height=2.0in,
  angle=-90]{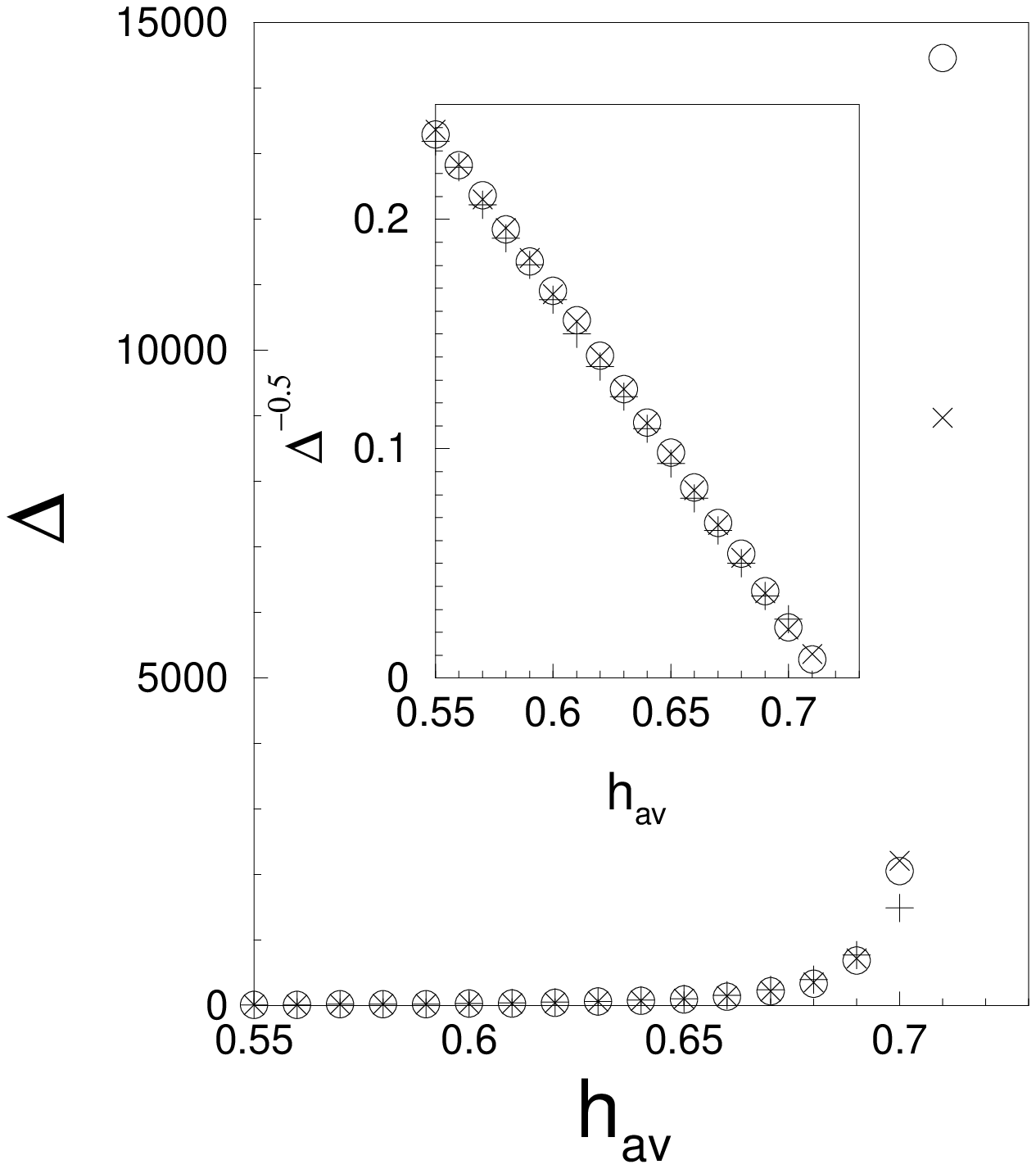}\hskip.2in\includegraphics[%
  width=2.5in,
  height=2.0in,
  angle=-90]{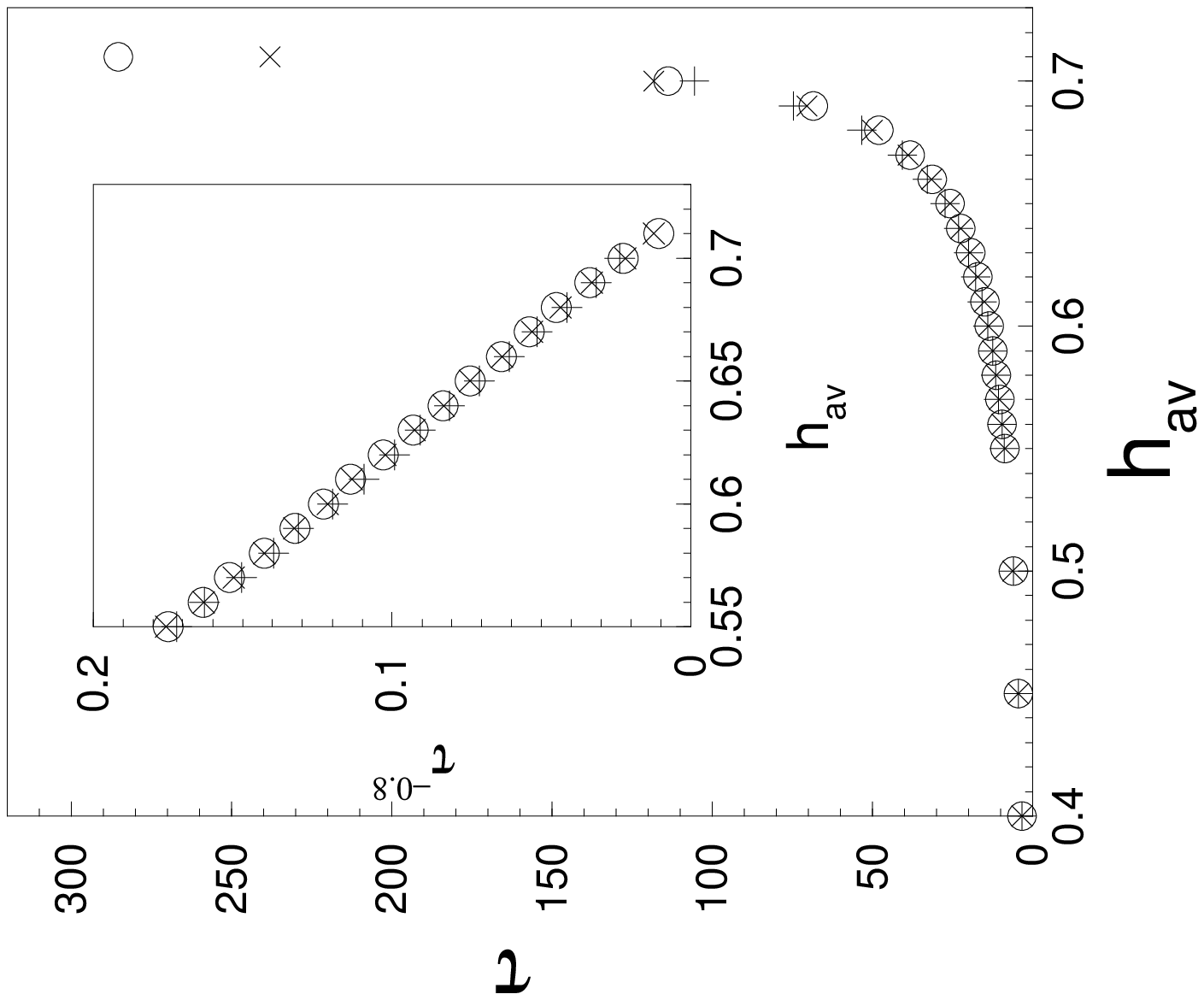} }\end{center}

\begin{center}{\includegraphics[%
  width=2.5in,
  height=2.5in,
  angle=-90]{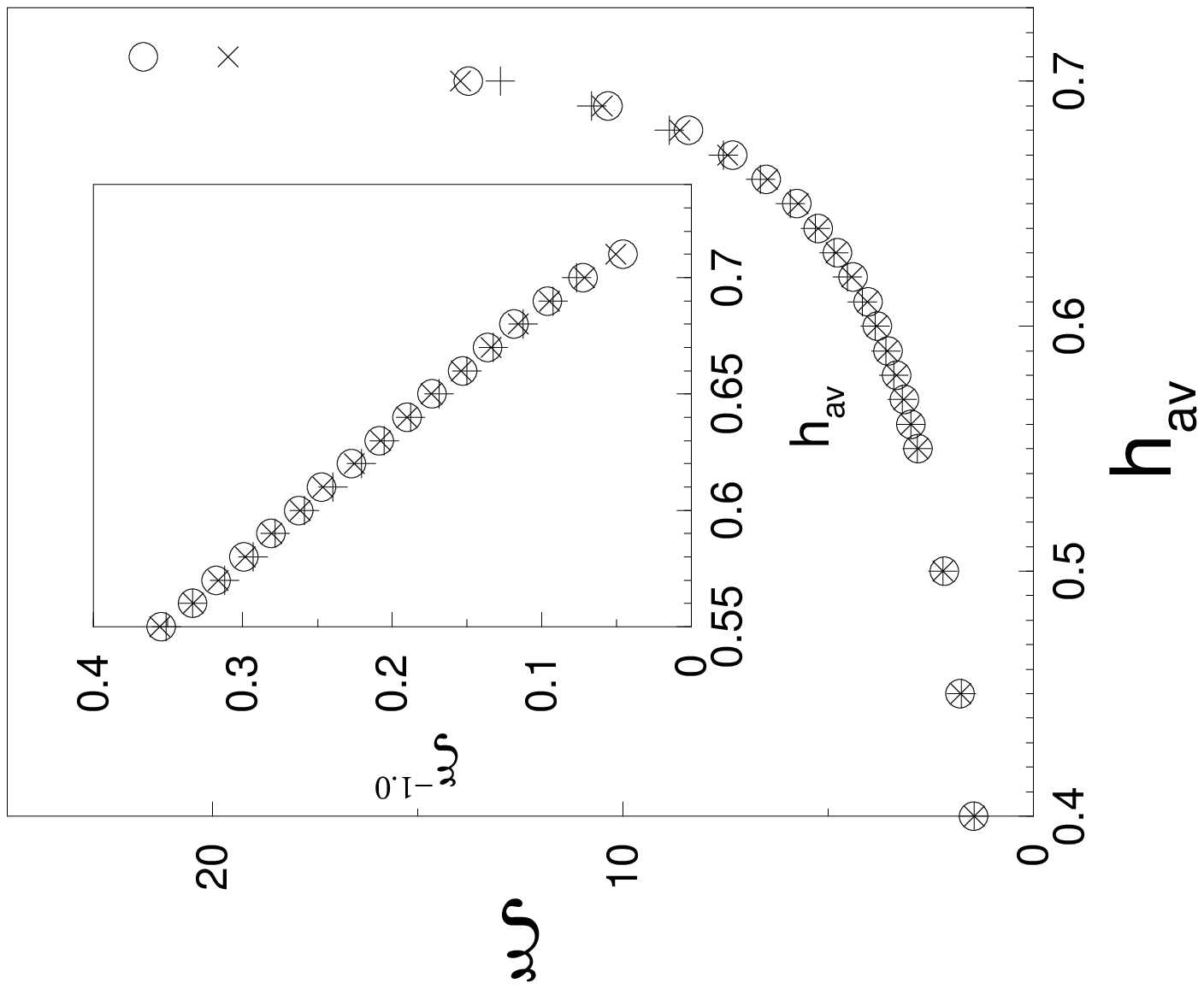}}\end{center}

{\small Fig. 14: Precursor parameters and prediction
of critical point in Manna model.}{\small \par}

\vskip.1in

Our simulation results suggest that although BTW and Manna models
belong to different universality class\index{universality class} with respect to their properties
at the critical state, both the models show similar sub-critical response
or precursors\index{precursor}. A proper extrapolation method can estimate the respective
critical heights of the models quite accurately.

\subsection{\noindent Fractal overlap model of earthquake }

\noindent It has been claimed recently that since the fractured surfaces
have got well-characterized self-affine\index{self-affine} properties, the distribution
of the elastic energies released during the slips between two fractal
surfaces (earthquake events) may follow the overlap distribution of
two self-similar fractal surfaces \cite{pc-Guten,pc-BK67,pc-CL89,pc-V96}. To
support this idea, Chakrabarti and Stinchcombe\index{Chakrabarti-Stinchcombe 
model} \cite{pc-BS99} have analytically
shown using renormalization group technique that for regular fractal\index{fractal}
overlap (Cantor sets and carpets) the contact area distribution follows
power law. This claim has also been verified by extensive numerical
simulations \cite{pc-PCRD-03}. If one cantor set moves uniformly over
other, the overlap between the two fractals (Fig. 15) change quasi-randomly
with time. In this section we analyse the time series data of such
overlaps to find the prediction possibility of a next large overlap.

\begin{center}\includegraphics[%
  width=5cm,
  height=3cm]{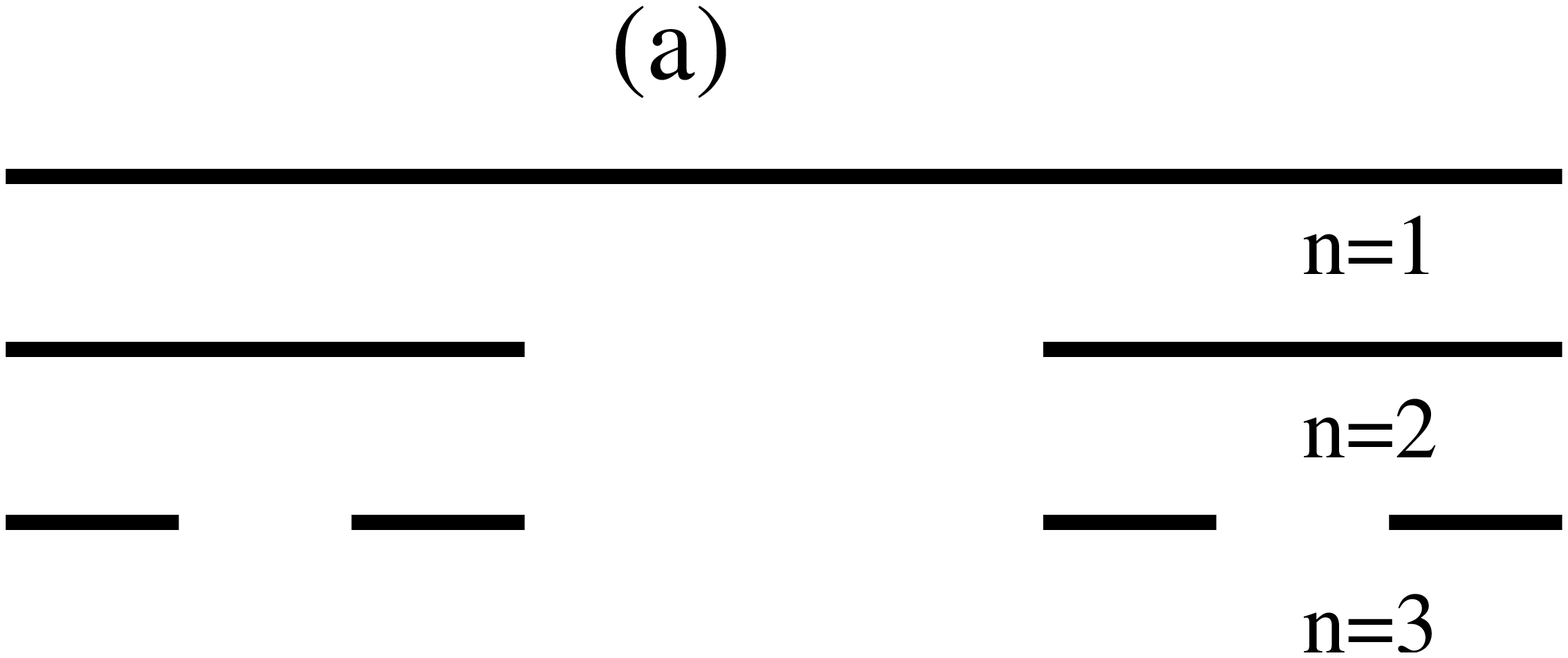}\hskip.5in\includegraphics[%
  width=5cm,
  height=3cm]{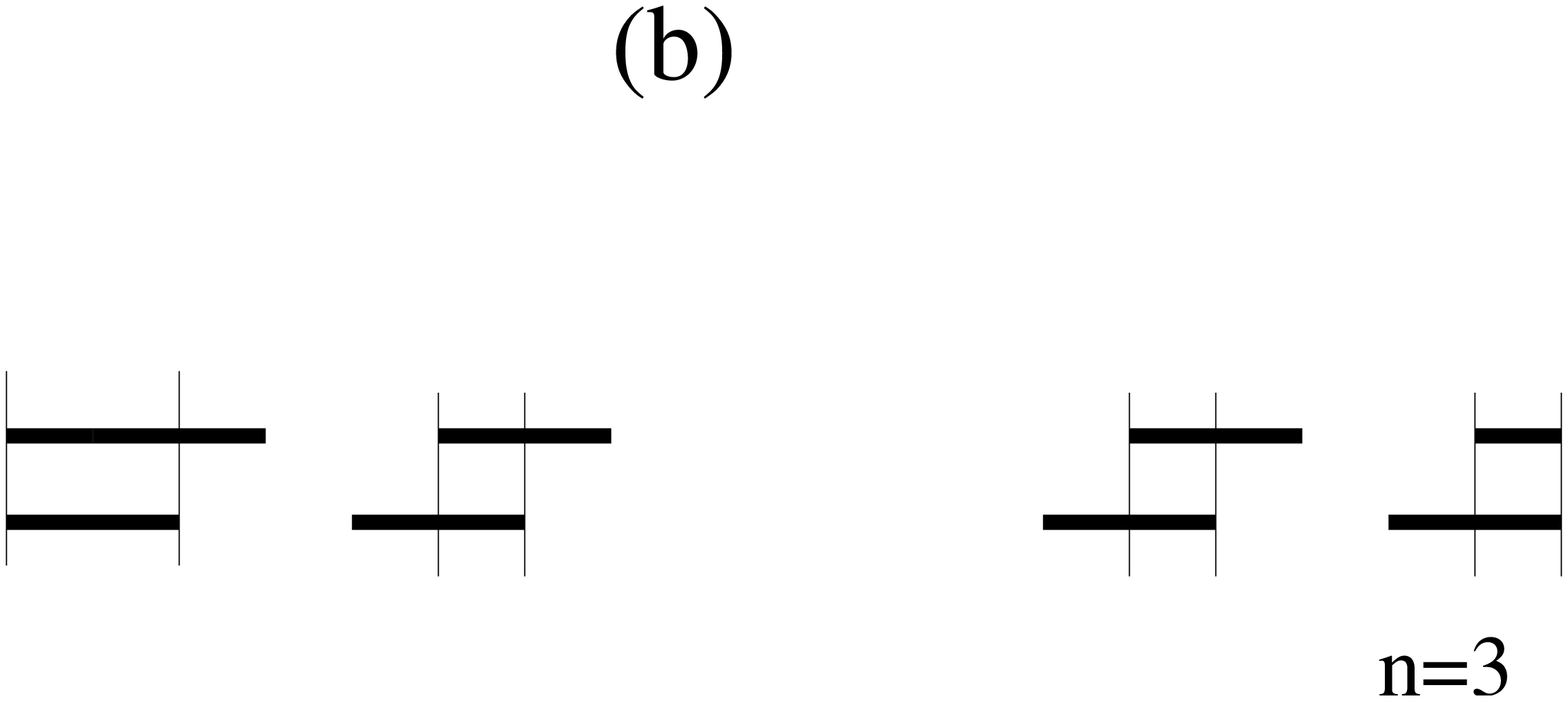}\end{center}

\vskip.1in

{\footnotesize Fig. 15:} \textbf{\footnotesize }{\footnotesize (a)
A regular Cantor set of dimension $\ln2/\ln3$; only three finite
generations are shown. (b) The overlap of two identical (regular)
Cantor sets, at $n=3$, when one slips over other; the overlap sets
are indicated within the vertical lines, where periodic boundary condition
has been used. }{\footnotesize \par}

\subsubsection{\noindent The time series data analysis }

\noindent We consider now the time series\index{time-series} obtained by counting the
overlaps $m(t)$ as a function of time as one Cantor set\index{Cantor set} moves over
the other (periodic boundary condition is assumed) 
with uniform velocity. The time series are shown
in Fig. 16., for finite generations of Cantor sets of dimensions $\ln2/\ln3$
and $\ln4/\ln5$ respectively.

We calculate the cumulative overlap size $Q(t)=\int_{o}^{t}mdt$.
We see that `on average' $Q(t)$ is seen to grow linearly with time
$t$ for regular Cantor sets. This gives a clue that instead of looking
at the individual overlaps $m(t)$ series one may look for the cumulative
quantity. In fact, for the regular Cantor set of dimension $\ln2/\ln3$,
the overlap $m$ is always $2^{k}$, where $k$ is an integer. However
the cumulative $Q(t)=\sum_{i=0}^{t}2^{k_{i}}$ can not be easily expressed
as any simple function of $t$. Still, we observe $Q(t)\simeq ct$,
where $c$ is dependent on $M$.

\includegraphics[%
  width=5cm,
  height=5cm]{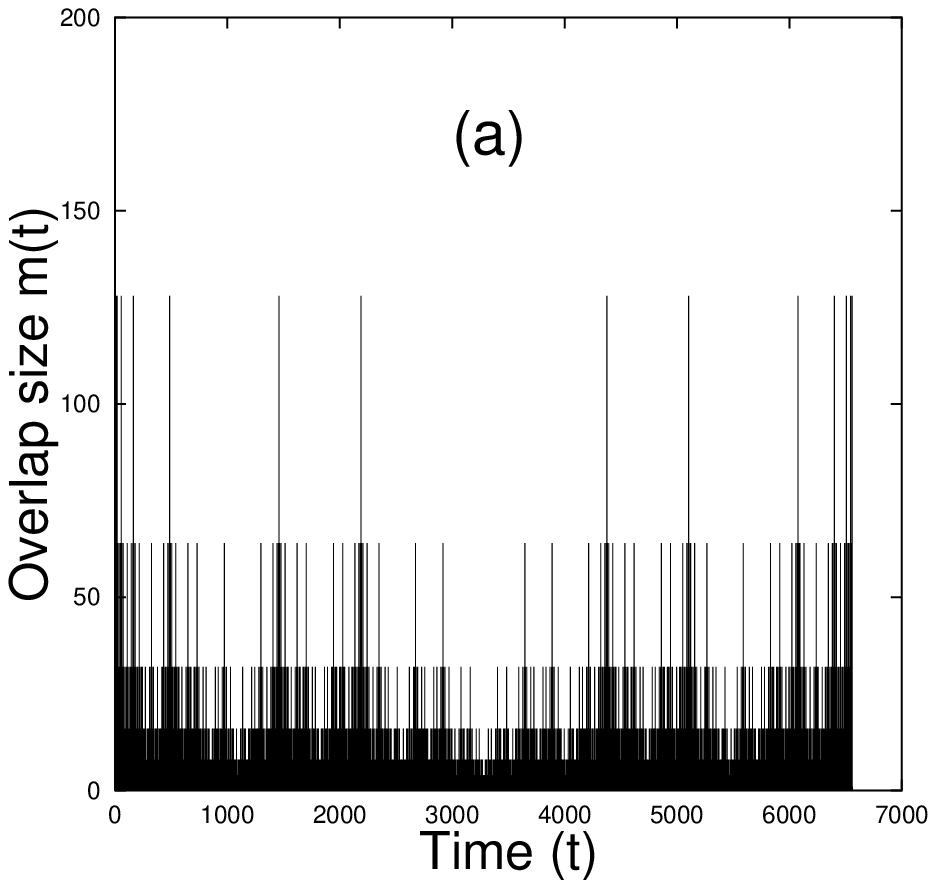}\hskip.2in\includegraphics[%
  width=5cm,
  height=5cm]{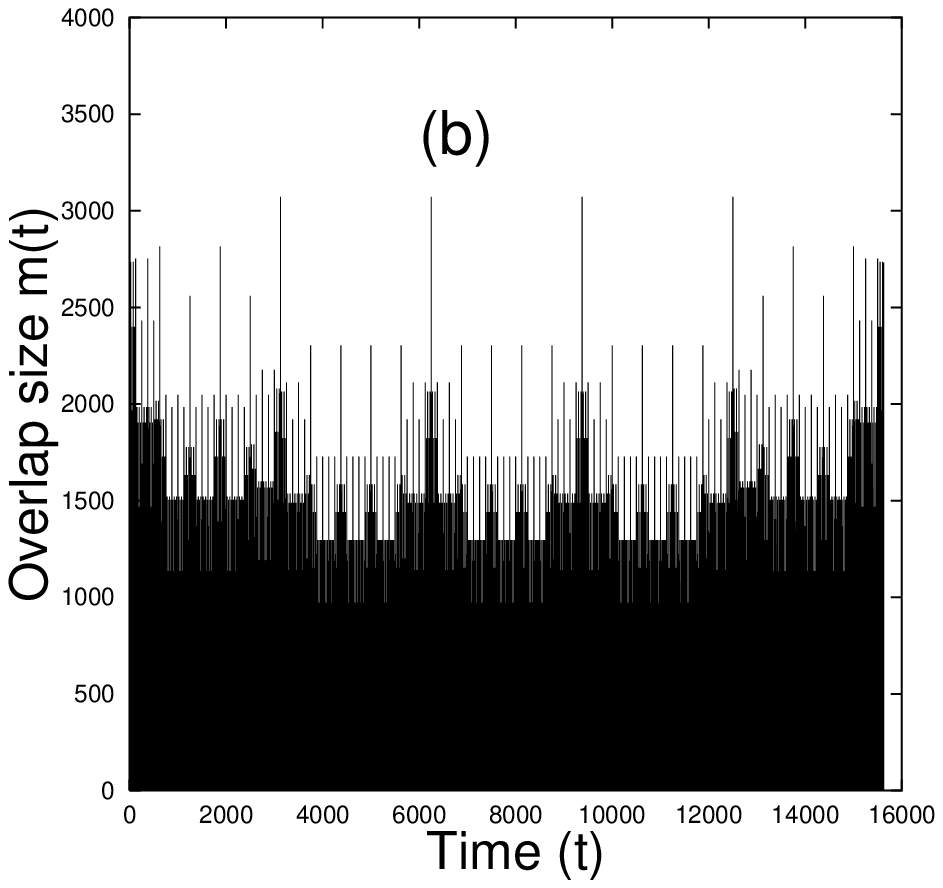}

\vskip.1in

{\small Fig. 16:} \textbf{\small }{\small The time ($t$) series data
of overlap size ($m$) for regular Cantor sets: (a) of dimension $\ln2/\ln3$,
at $8$th generation: (b) of dimension $\ln4/\ln5$, at $6$th generation.}{\small \par}

\vskip.1in

We first identify the `large events' occurring at time $t_{i}$ in
the $m(t)$ series, where $m(t_{i})\geq M$, a pre-assigned number.
Then, we look for the cumulative overlap\index{cumulative overlap} size $Q(t)=\int_{t_{i}}^{t_{i+1}}mdt$,
where the successive large events occur at times $t_{i}$ and $t_{i+1}$.
The behavior of $Q_{i}$ with time is shown in Fig. 17 for regular
cantor sets with $d_{f}=$$\ln2/\ln3$ at generation $n=8$. Similar
results are also given for Cantor sets with $d_{f}=$$\ln4/\ln5$
at generation $n=6$ in Fig. 18. It appears that there are discrete
(quantum) values  up to which $Q_{i}$ grows with time
$t$ and then drops to zero value.

\includegraphics[%
  width=5cm,
  height=5cm]{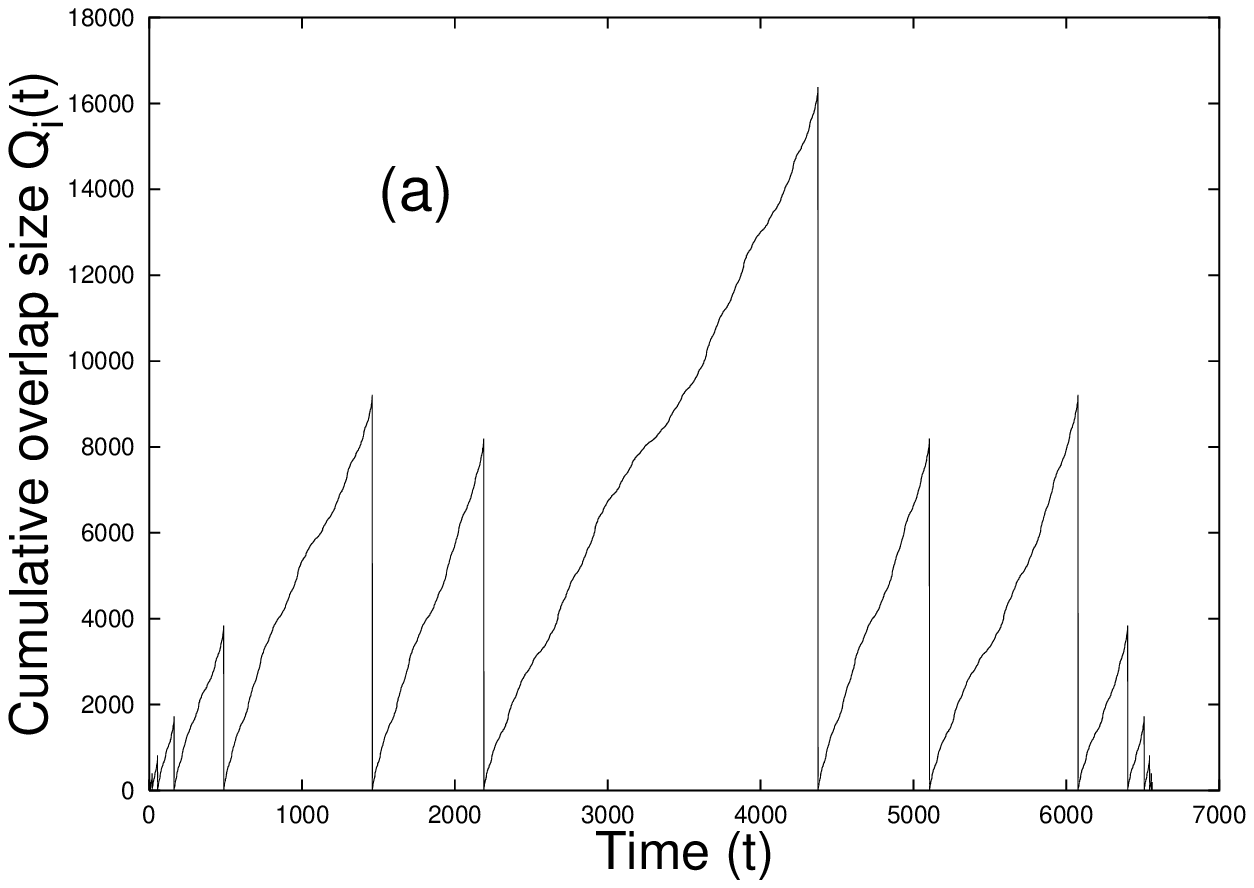}\hskip.3in\includegraphics[%
  width=5cm,
  height=5cm]{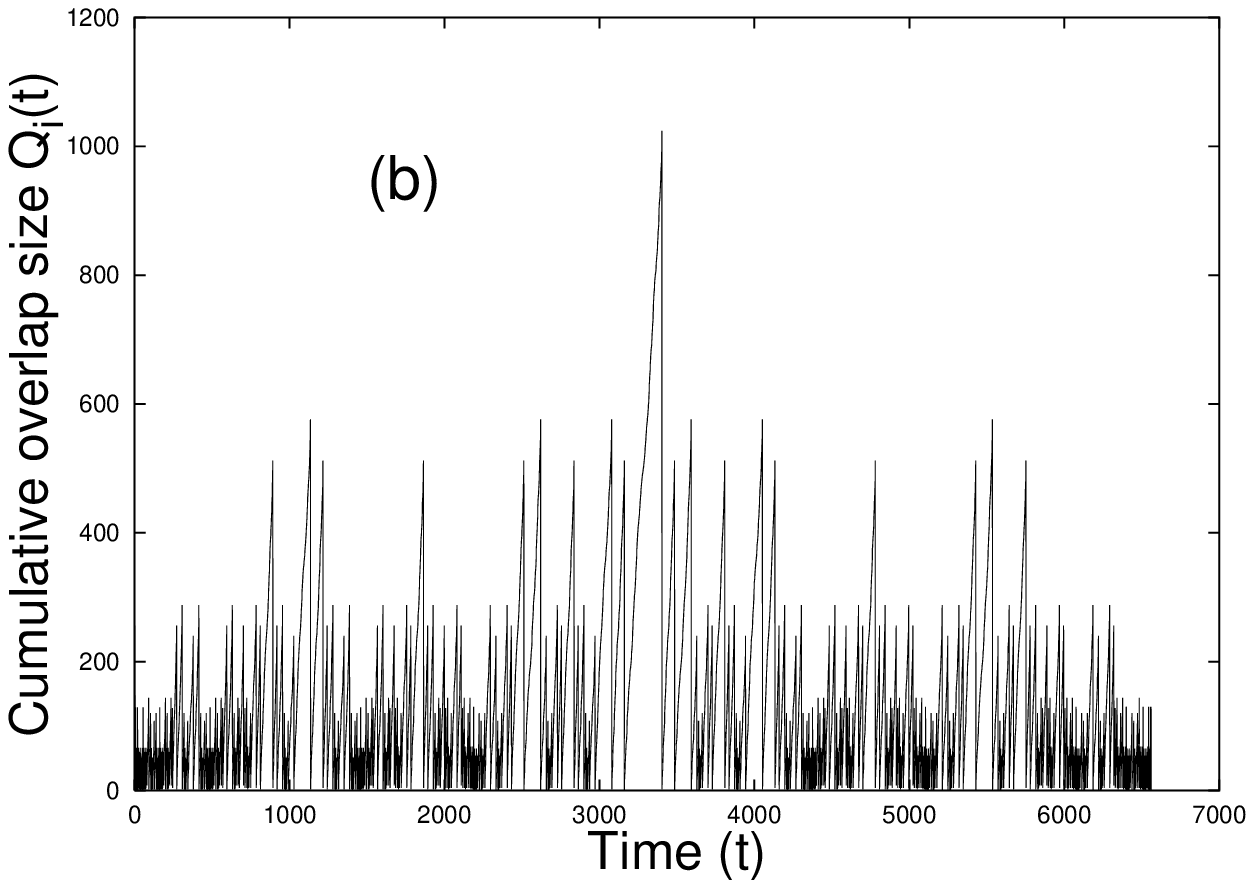}

\vskip.1in

{\small Fig. 17: The cumulative overlap size variation with time (for
regular Cantor sets of dimension $\ln2/\ln3$, at $8$th generation),
where the cumulative overlap has been reset to $0$ value after every
big event (of overlap size $\geq M$ where $M=128$ and $32$ respectively).} 

\includegraphics[%
  width=5cm,
  height=5cm]{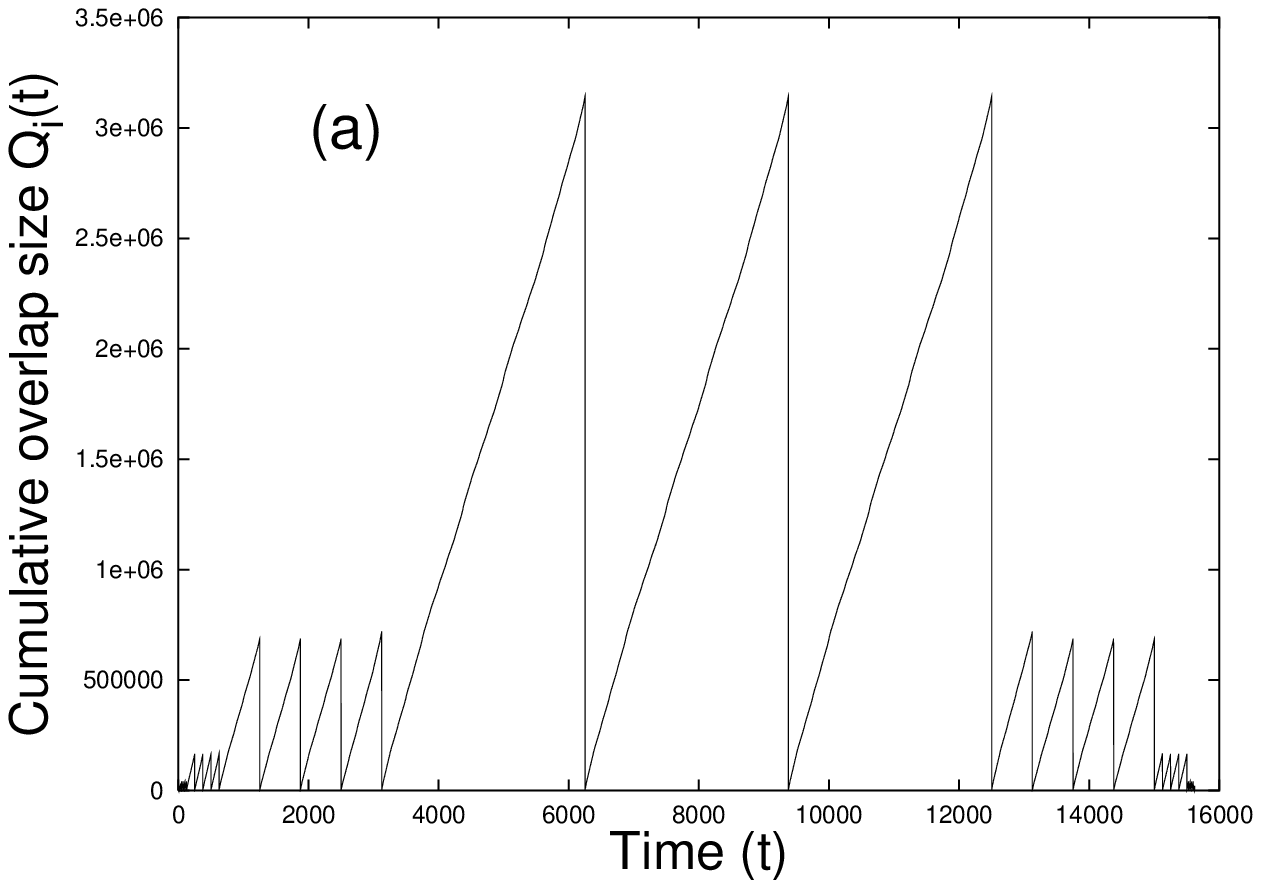}\hskip.3in\includegraphics[%
  width=5cm,
  height=5cm]{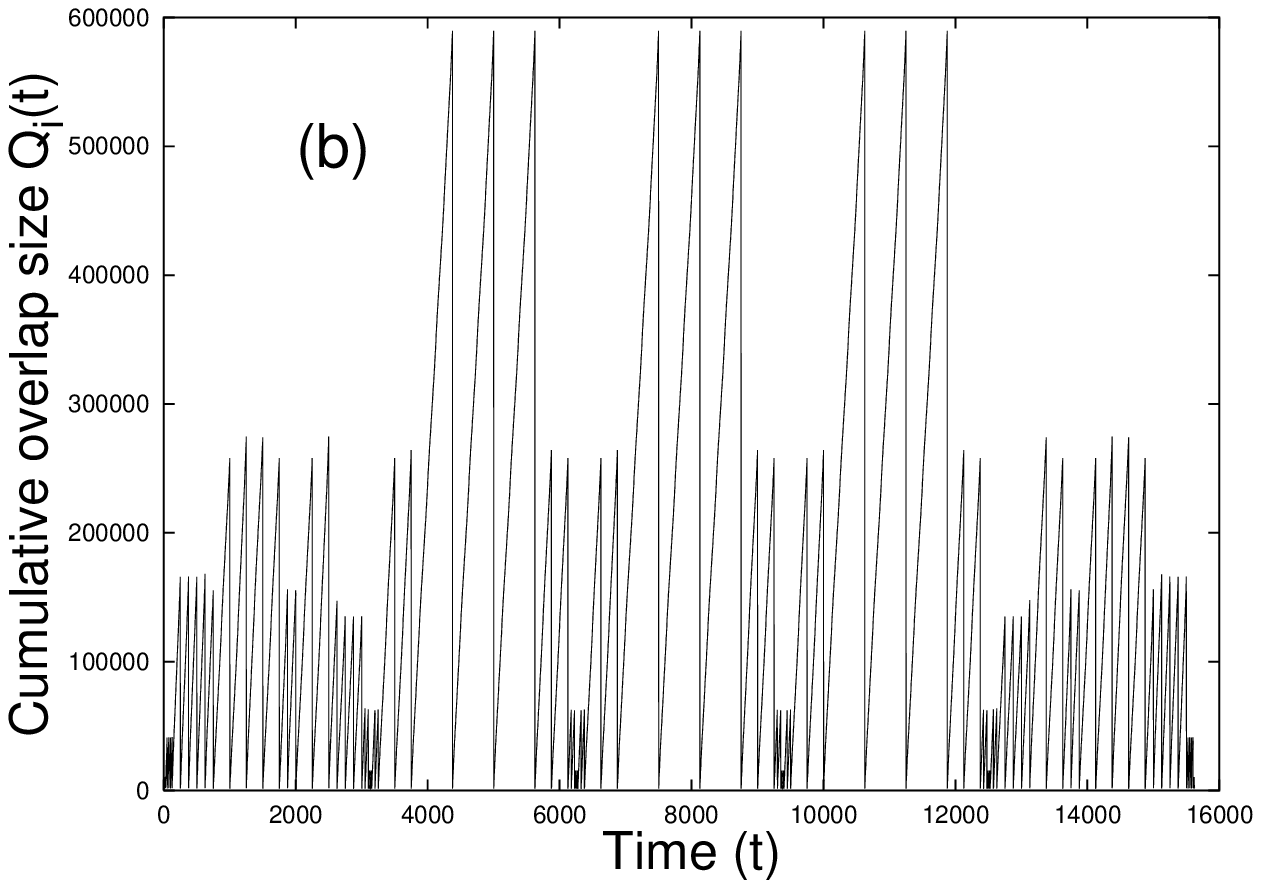}

{\small Fig. 18: The cumulative overlap size variation with time (for
regular Cantor sets of dimension $\ln4/\ln5$, at $6$th generation),
where the cumulative overlap has been reset to $0$ value after every
big event (of overlap size $\geq M$ where $M=2400$ and $2048$ respectively). }{\small \par}

\vskip.1in

Finally we found that if one fixes a magnitude $M$ of the overlap
sizes $m$, so that overlaps with $m\geq M$ are called `events'\index{earthquake} (or
earthquake), then the cumulative overlap $Q_{i}$ grows linearly with
time up to some discrete quanta $Q_{i}\cong lQ_{0}$, where $Q_{0}$
is the minimal overlap quantum, dependent on $M$ and $l$ is an integer.

\section{\noindent Conclusions}

Knowledge of precursors sometimes help to estimate precisely the
location of the global failure or critical point through a proper
extrapolation procedure. Therefore precursors which are available
long before the global failure, can be used to resist an imminent
global failure. 

In all the dynamical systems studied here, we find that long before
the occurrence of global failure, the growing correlations in the
dynamics of constituent elements manifest themselves as various precursors.
In  Fiber Bundle Model,  the breakdown susceptibility
$\chi$ and the relaxation time $\tau$, both diverge
as the external load or stress approaches the global failure point
or critical point. The distribution of avalanches exhibits a crossover in
 power law exponent values when the system comes closer to the failure point.  
Also the pattern 
of inclusive avalanches can tell us whether a bundle can support the applied 
load or not. In Fuse model, divergence of susceptibility and 
crossover in avalanche power law exponent are the signature of imminent 
breakdown of the system.  Whereas in both BTW
and Manna sandpile models the number of toppling $\Delta$, relaxation
time $\tau$ and the correlation length $\xi$ grow and diverge following
power laws as the systems approach their respective critical points
$h_{c}$ from the sub-critical states. Therefore these parameters directly 
help to predict the critical point in advance. However in 
fractal overlap model the time series data analysis suggest that the 
cumulative overlap sizes can assume some quantized values which indirectly 
helps to speculate whether a large overlap (event) is imminent or not. 

\vskip.3in
\textbf{Acknowledgment:} S. P. thanks the Research Council of Norway (NFR) for
 financial support through Grant No. 166720/V30.

\printindex
\end{document}